\documentclass[12pt]{amsart}
\usepackage{amsmath,amssymb,enumitem,hyperref,cite}

\hypersetup{pdftex,colorlinks=true,allcolors=blue}
\usepackage[all]{hypcap}

\usepackage[foot]{amsaddr}
\usepackage[sc]{mathpazo}

\newcommand\ack{\subsection*{Acknowledgment}}

\DeclareMathAlphabet\mathsfbi{T1}{phv}{b}{it}

\numberwithin{equation}{section}

\newcommand\BV{\boldsymbol} 
\newcommand\BM{\mathsfbi} 

\newcommand\dif{\,\mathrm{d}}
\newcommand\deriv[2]{\frac{\mathrm{d} #1}{\mathrm{d} #2}}
\newcommand\parderiv[2]{\frac{\partial #1}{\partial #2}}
\newcommand\Dderiv[2]{\frac{\mathrm{D} #1}{\mathrm{D} #2}}
\newcommand\RR{\mathbb R}

\newcommand\Mach{\mathit{Ma}}
\newcommand\Rey{\mathit{Re}}
\DeclareMathOperator\trace{tr}
\DeclareMathOperator\diag{diag}

\newcommand\hE{\,\widehat{\!\BM E}}

\def \HS {\mathrm{\scriptscriptstyle HS}}
\def \NS {\mathrm{\scriptscriptstyle NS}}
\def \odeVec {{\BV X}}

\begin{document}

\author[Rafail V. Abramov]{Rafail V. Abramov}

\address{Department of Mathematics, Statistics and Computer Science,
University of Illinois at Chicago, 851 S. Morgan st., Chicago, IL 60607}

\email{abramov@uic.edu}

\title[Turbulence via intermolecular potential: Uncovering the origin]
{Turbulence via intermolecular potential:\\ Uncovering the origin}

\begin{abstract}
In recent works, we proposed a hypothesis, according to which
turbulence in gases is created by the mean field effect of an
intermolecular potential. We discovered that, in a numerically
simulated inertial flow, turbulent solutions indeed spontaneously
emerge from a laminar initial condition, as observed in nature and
experiments. To study the origin of turbulent dynamics, in the current
work we examine the equations of a two-dimensional inertial flow,
linearized around a large scale constant vorticity state. Remarkably,
even in this simplified setting, we find that turbulent dynamics
emerge as linearly unstable fluctuations of the velocity divergence.

In particular, for the linearized dynamics at a high Reynolds number,
we find that, at short time scales, the coupling of the mean field
potential with the large scale background vorticity creates linearly
unstable, rapidly oscillating fluctuations of the divergence of
velocity at inertial scales. In the asymptotic time limit, we find a
persistent eigenvector, also aligned largely with the divergence of
velocity, which allows these fluctuations to propagate in the form of
traveling waves in the Fourier domain. Remarkably, these traveling
waves decay at a constant, scale-independent exponential rate, which
is rather unusual, because all explicitly dissipative terms in the
equations are viscous. Furthermore, it appears that the famous
Kolmogorov scaling of the kinetic energy is produced by this
persistent velocity divergence, due to a cubic relation between the
physical time variable, and a pseudo-time variable, in which the
dynamics become asymptotically autonomous. These effects vanish when
the mean field potential is removed.
\end{abstract}

\maketitle

\section{Introduction}

In his famous experiment, Reynolds~\cite{Rey83} discovered that an
initially laminar flow of water in a straight smooth pipe
spontaneously develops turbulent motions whenever the high Reynolds
number condition is satisfied, even if all reasonable care is taken to
avoid disturbing the flow artificially. Subsequently, numerous
attempts to explain the physical nature of turbulence have been made
throughout the twentieth century (see~\cite{Cho,Fri} for a detailed
exposition), yet, the reason for such a spontaneous manifestation of
turbulence in an initially laminar flow remained unknown thus far.

Conventionally, fluid flows at low Mach numbers are modeled via the
incompressible Navier--Stokes equations~\cite{MajBer}:
\begin{equation}
\label{eq:NS}
\rho_0\Dderiv{\BV u}t=-\nabla p+\mu\Delta\BV u,\qquad\nabla\cdot\BV
u=0.
\end{equation}
Above, $\rho_0$ is the constant density of the flow, $\BV u$ is its
velocity, $p$ is pressure, and $\mu$ is dynamic viscosity (which is
treated as a constant parameter, for simplicity). The symbol $\mathrm
D/\mathrm Dt$ denotes the advection operator (or, as it is also known,
the ``material derivative''):
\begin{equation}
\Dderiv{}t\equiv\parderiv{}t+\BV u\cdot\nabla.
\end{equation}
It is known that laminar solutions of \eqref{eq:NS} do not
spontaneously become turbulent; in fact, in numerically simulated
solutions of~\eqref{eq:NS}, turbulent-like motions are created via
deliberate perturbations
\cite{AviMoxLozAviBarHof,BarSonMukLemAviHof,KhaAnwHasSan,Vel}. Such a
notorious inability of the incompressible Navier--Stokes equations to
produce spontaneously turbulent solutions suggests that a key physical
effect, which is responsible for the development of turbulence, is
missing from them.

In our recent works \cite{Abr22,Abr23,Abr24,Abr26} we proposed and
numerically investigated a hypothesis, according to which turbulence
in an initially laminar, inertial (that is, constant pressure) gas
flow is created by the mean field effect of molecules interacting
through their intermolecular potential. The system of transport
equations for such an inertial flow is
\begin{equation}
\label{eq:inertial_flow}
\Dderiv\rho t=-\rho\nabla\cdot\BV u,\qquad\rho\Dderiv{\BV u}t=-\nabla
\bar\phi+\mu\Delta\BV u,
\end{equation}
where $\bar\phi$ is the corresponding mean field potential, while the
density $\rho$ is no longer constant and has its own transport
equation. Under the assumption that the gas is sufficiently dilute,
and that its molecules can be approximated by hard spheres of the
appropriate mass and diameter, the mean field potential $\bar\phi$ is
given by the following formula \cite{Abr22,Abr23,Abr24,Abr25,Abr26}:
\begin{equation}
\label{eq:bphi}
\bar\phi=\frac{4\rho p}{\rho_\HS}.
\end{equation}
Above, $\rho_\HS$ is the mass density of a hard sphere.

The term $\nabla\bar\phi$ is not present in the conventional equations
of fluid mechanics, because, in the Boltzmann--Grad limit \cite{Gra},
$\rho_\HS\to\infty$ and $\bar\phi$ subsequently vanishes. However, we
discovered computationally that, in our model of inertial gas flow,
turbulent motions spontaneously emerge in the presence of $\bar\phi$,
while failing to do so in its absence~\cite{Abr22,Abr23,Abr24,Abr26}.
Moreover, we have shown \cite{Abr25} that, at low Mach numbers,
$\nabla\bar\phi\sim 1$ in the corresponding nondimensional variables,
which means that the effect of $\bar\phi$ is non-negligible.

We have to note that, in nature, purely inertial flows are observed
under rather specific conditions -- for example, a quasigeostrophic
flow \cite{Cha} on a planetary scale near the equator, where the
Coriolis effect is weak. However, unlike the conventional compressible
and incompressible models, the inertial flow model
\eqref{eq:inertial_flow} reproduces the key thermodynamic property of
a gas flow at a low Mach number, which is the tendency to expand when
the temperature increases. In particular, in the presence of gravity,
this property becomes the primary cause of dry convection in the
atmosphere.  Conversely, neither the compressible nor incompressible
Navier--Stokes equations possess this property -- the density is
constant in the latter, whereas in the former the gas compresses when
heated due to the entropy becoming invariant for a small
viscosity~\cite{Abr24}.

Having ascertained computationally that the presence of the mean field
potential correlates with the spontaneous manifestation of turbulent
motions in a numerically simulated inertial flow, in the current work
we conduct an elementary linear analysis of the inertial flow
equations \eqref{eq:inertial_flow} in a two-dimensional setting, and
compare it to the similarly linearized incompressible Navier--Stokes
equations \eqref{eq:NS}. The motivation behind this work is the
following. Since we found, via numerical simulations, that turbulent
motions develop from a laminar background steady state spontaneously,
it means that they originate from small fluctuations around the
background state. In nature, these small fluctuations are provided by
various processes not accounted for in the equations (thermal noise,
randomly passing acoustic waves, etc), while in numerical simulations
the same role is fulfilled by the machine round-off errors. Therefore,
there exists a stage of development of turbulent motions where the
fluctuations are still sufficiently small, so that they can be
described by the linearized dynamics. Additionally, at this stage of
flow development, the large scale background velocity can be
approximated by a linear expression, which effectively confines the
flow to a plane, and sets its background vorticity to a constant.
Thus, in order to study this stage, it suffices to examine a
two-dimensional flow linearized around a constant vorticity state.

Remarkably, we find that the coupling of the mean field potential with
the large scale vorticity of the flow induces rapid oscillations at
short time scales in small-scale, linearly unstable fluctuations of
the divergence of velocity, and the frequency of these oscillations is
inversely proportional to the spatial scale of the fluctuation.
Conversely, in the absence of $\bar\phi$, we find that, although some
oscillations still occur, their frequency is slow across all spatial
scales.  At long time scales, we find that there exists an
eigenvector, which is also aligned largely with the divergence of
velocity, and which allows persistent solutions to propagate along
characteristics in the Fourier domain in the form of traveling waves,
which uniformly decay at a constant exponential rate.  Further, it
appears that these waves of the divergence of velocity produce the
famous Kolmogorov energy scaling of the negative $5/3$-power of the
wavenumber \cite{Kol41a,Kol41c}. This peculiar power of the scaling
arises due to the cubic time dilation in asymptotically autonomous
solutions along characteristics, which is a fundamental property of
the dynamics in \eqref{eq:inertial_flow}. At the same time, generic
solutions of the similarly linearized incompressible Navier--Stokes
equations~\eqref{eq:NS} decay monotonically to zero.

The work is organized as follows. In Section~\ref{sec:2D_flow} we show
that a typical prototype scenario with laminar background flow and
large scale vorticity can be formally confined to a plane (with the
third dimension being neutral). We also use the Helmholtz
decomposition to reformulate the two-dimensional inertial flow
equations in terms of vorticity and divergence of velocity. In
Section~\ref{sec:linearization} we linearize the vorticity-divergence
equations around a large scale constant vorticity state. In
Section~\ref{sec:short_time} we show that, at short time scales and
linearly unstable wavenumbers, the coupling of the mean field
potential with large scale vorticity induces rapid oscillations of the
divergence of velocity.  In Section~\ref{sec:long_time} we show that,
although the linearized system is asymptotically stable along
characteristics, there is an eigenvector, aligned largely with the
divergence of velocity, which allows persistent solutions to propagate
along characteristics as traveling waves decaying at a constant,
scale-independent exponential rate. In Section~\ref{sec:power_scaling}
we estimate the Kolmogorov scaling of the kinetic energy by making use
of the cubic dependence between the physical time variable, and a
rescaled pseudo-time variable, which causes the solutions to become
autonomous along characteristics. In Section~\ref{sec:discussion} we
discuss the results.

\section{The equations for inertial flow in a plane}
\label{sec:2D_flow}

As a starting point, we need to convert the inertial flow equations in
\eqref{eq:inertial_flow} into a form which is sufficiently simple and
convenient for subsequent analysis. To this end, we first show that,
in a basic ``prototype'' flow scenario, where spontaneous
manifestation of turbulence is usually observed, the reference frame
can be chosen so that the third dimension becomes neutral, and
\eqref{eq:inertial_flow} can be confined to a plane. In this plane, we
subsequently take advantage of Helmholtz's decomposition, and
reformulate the equations in terms of the vorticity and velocity
divergence variables.

\subsection{The prototype laminar background flow}
\label{sec:background_flow}

Based on observations, the simplest scenario, where turbulent motions
develop spontaneously, is a steady large scale laminar flow of
constant density in a fixed direction, whose speed varies in the plane
orthogonal to the direction of the flow. In the Cartesian reference
frame, without loss of generality, we can rotate the coordinate axes
so that the $x$-axis is parallel to the direction of the flow. In such
a reference frame, the large scale background velocity field $\BV
u_0(\BV x)$ is of the form
\begin{equation}
\BV u_0(\BV x)=(U(y,z),\,0,\,0),
\end{equation}
where $U(y,z)$ is the speed of the large scale flow. Since $U(y,z)$ is
presumed to be a slowly-varying function on the spatial scale of
interest, let us expand it in Taylor series around a reference point,
which we take to be zero without loss of generality:
\begin{equation}
U(y,z)=U(0,0)+\nabla U(0,0)\cdot(y,\, z)+o(\|(y,\, z)\|).
\end{equation}
First, using the Galilean shift of the reference frame along its
$x$-axis, we can eliminate the constant term $U(0,0)$. Second, we can
rotate the $yz$-plane around the $x$-axis so that the direction of
$\nabla U$ is aligned with the direction of the $y$-axis. As a result,
$U(y,z)$ acquires the form
\begin{equation}
U(y,z)=\left.\parderiv Uy\right|_{(0,0)}y+o(\|(y,\, z)\|).
\end{equation}
Next, we define the large scale vorticity $\Omega$ via
\begin{equation}
\Omega=\left.-\parderiv Uy\right|_{(0,0)}>0,
\end{equation}
where the positive sign is chosen without loss of generality. Therefore,
\begin{equation}
U(y,z)=-\Omega y+o(\|(y,\, z)\|),\qquad\BV u_0(\BV x)=(-\Omega y,\, 0,
\,0)+o(\|(y,\, z)\|).
\end{equation}
According to observations, turbulent motions manifest across a broad
range of scales, from centimeters to hundreds of kilometers, while the
large scale background flow can vary quite slowly, across hundreds of
kilometers (for example, the Earth atmosphere). Therefore, we can
assume that the spatial scale of the problem is much smaller than the
scale of the curvature of $\BV u_0(\BV x)$, and keep only the leading
order term of the Taylor series:
\begin{equation}
\label{eq:linear_shear}
\BV u_0(\BV x)=(-\Omega y,\, 0,\, 0).
\end{equation}
Here, the $z$-direction is completely neutral, and, therefore, the
problem is effectively two-dimensional. Additionally, in our recent
work \cite{Abr23}, we found that turbulent motions spontaneously
manifest in a numerically simulated two-dimensional inertial flow.
This suggests that, at least at the initial stage of its development,
turbulence is effectively a two-dimensional phenomenon.

\subsection{Change of variables and Helmholtz's decomposition}

With \eqref{eq:bphi}, the equations for inertial flow in
\eqref{eq:inertial_flow} can be written as
\begin{equation}
\label{eq:inertial_flow_2}
\Dderiv\rho t=-\rho\nabla\cdot\BV u,\qquad\rho\Dderiv{\BV u}t=-\frac{4
  p_0}{\rho_\HS}\nabla\rho+\mu\Delta\BV u,
\end{equation}
where $p_0$ is the constant background pressure.  As a first step, we
replace $\rho=\rho_0e^\zeta$, where $\rho_0$ is a reference
density. This change of variable transforms \eqref{eq:inertial_flow_2}
into
\begin{equation}
\label{eq:inertial_flow_zeta}
\Dderiv\zeta t=-\nabla\cdot\BV u,\qquad\Dderiv{\BV u}t=-\frac{4p_0
\eta}{\rho_0}\nabla\zeta+\nu e^{-\zeta}\Delta\BV u,
\end{equation}
where $\nu$ and $\eta$ are the kinematic viscosity and the packing
fraction, respectively, given via
\begin{equation}
\nu=\frac\mu{\rho_0},\qquad\eta=\frac{\rho_0}{\rho_\HS}.
\end{equation}
Above, we established that, without much loss of generality, a
prototype problem with large scale laminar background flow can be
confined to a plane by means of an appropriate choice of the frame of
reference. According to Helmholtz's decomposition, in a plane, a
vector field $\BV u$ can be decomposed into the stream function and
the potential:
\begin{equation}
\BV u=\nabla^\perp\psi+\nabla\phi,\qquad\nabla=\begin{pmatrix}
\partial/\partial x \\ \partial/\partial y\end{pmatrix},\qquad
\nabla^\perp=\begin{pmatrix}-\partial/\partial y \\ \partial /\partial
x\end{pmatrix}.
\end{equation}
Above, $\psi$ is the stream function, and $\phi$ is the potential.
Below, we denote the vorticity and divergence of $\BV u$,
respectively, via
\begin{equation}
\omega=\nabla^\perp\cdot\BV u=\Delta\psi,\qquad\chi=\nabla\cdot\BV u=
\Delta\phi.
\end{equation}
Applying the divergence and curl to the velocity equation in
\eqref{eq:inertial_flow_zeta}, we arrive at
\begin{equation}
\nabla\cdot\Dderiv{\BV u}t=-\frac{4p_0\eta}{\rho_0}\Delta\zeta +\nu
\nabla\cdot (e^{-\zeta}\Delta\BV u),\qquad\nabla^\perp\cdot\Dderiv{\BV
  u}t=\nu \nabla^\perp\cdot(e^{-\zeta}\Delta\BV u).
\end{equation}
Next, we observe that
\begin{subequations}
\begin{equation}
\nabla\cdot\Dderiv{\BV u}t=\parderiv{(\nabla\cdot\BV u)}t+(\BV u\cdot
\nabla)(\nabla\cdot\BV u)+\nabla\BV u^T:\nabla\BV u=\Dderiv\chi t+
\|\nabla(\nabla^\perp\psi+\nabla\phi)\|_F^2-\omega^2,
\end{equation}
\begin{equation}
\nabla^\perp\cdot\Dderiv{\BV u}t=\parderiv{(\nabla^\perp\cdot\BV u)}t
+(\BV u\cdot\nabla)(\nabla^\perp\cdot\BV u)+\nabla^\perp\BV u^T:\nabla
\BV u=\Dderiv\omega t+\omega\chi,
\end{equation}
\begin{equation}
\nabla\cdot (e^{-\zeta}\Delta\BV u)=e^{-\zeta}\big(\Delta(\nabla\cdot
\BV u)-\nabla \zeta\cdot\Delta\BV u\big)=e^{-\zeta}\big(\Delta\chi
-\nabla\zeta\cdot(\nabla^\perp\omega+\nabla\chi)\big),
\end{equation}
\begin{equation}
\nabla^\perp\cdot (e^{-\zeta}\Delta\BV u)=e^{-\zeta}\big(\Delta(\nabla
^\perp\cdot\BV u)-\nabla^\perp\zeta\cdot\Delta\BV u\big)=e^{-\zeta}
\big(\Delta\omega -\nabla^\perp\zeta\cdot(\nabla^\perp\omega+\nabla
\chi)\big),
\end{equation}
\end{subequations}
where, $\|\cdot\|_F$ is the Frobenius norm of a matrix. This leads to
the following equations for $\zeta$, $\omega$ and $\chi$:
\begin{subequations}
\label{eq:vorticity_divergence}
\begin{equation}
\Dderiv\zeta t=-\chi,\qquad\Dderiv\omega t=-\omega\chi+\nu e^{-\zeta
}\big[\Delta\omega-\nabla^\perp\zeta\cdot(\nabla^\perp \omega+\nabla
  \chi)\big],
\end{equation}
\begin{equation}
\label{eq:divergence}
\Dderiv\chi t=\omega^2-\|\nabla(\nabla^\perp\psi+\nabla\phi)\|_F^2-
\frac{4p_0\eta}{\rho_0}\Delta\zeta+\nu e^{-\zeta}\big[\Delta\chi-\nabla\zeta
  \cdot(\nabla^\perp\omega+\nabla\chi)\big].
\end{equation}
\end{subequations}
At the same time, the incompressible Navier--Stokes equations in
\eqref{eq:NS} become a single vorticity equation after Helmholtz's
decomposition,
\begin{equation}
\label{eq:NS_vorticity}
\Dderiv{\omega_\NS}t=\nu\Delta\omega_\NS,
\end{equation}
as both $\zeta=0$ and $\chi=0$. For details, see, for example, Chapter
2 of~\cite{MajBer}.

\subsection{Nondimensionalization}
\label{sec:nondimensionalization}

It is useful to nondimensionalize the variables for further analysis.
In the current context, we assume that the spatial scale of the flow
is $L$, while the vorticity has a reference magnitude $\Omega$, taken
from the linear shear flow \eqref{eq:linear_shear}. We now introduce
the following rescalings:
\begin{subequations}
\begin{equation}
x=L\tilde x,\qquad y=L\tilde y,\qquad t=\Omega^{-1}\tilde t,
\end{equation}
\begin{equation}
\omega=\Omega\tilde\omega,\qquad\chi=\Omega\tilde\chi,\qquad
\psi=\Omega L^2\tilde\psi,\qquad\phi=\Omega L^2\tilde\phi.
\end{equation}
\end{subequations}
Here, $\zeta$ is already nondimensional. In the new variables, the
equations in \eqref{eq:vorticity_divergence} become
\begin{subequations}
\label{eq:vorticity_divergence_nondimensional}
\begin{equation}
\Dderiv\zeta{\tilde t}=-\tilde\chi,\qquad\Dderiv{\tilde\omega}{\tilde
  t}=-\tilde\omega\tilde\chi+\frac 1\Rey e^{-\zeta}\big[\tilde\Delta
  \tilde\omega-\tilde\nabla^\perp\zeta\cdot(\tilde\nabla^\perp\tilde
  \omega+\tilde\nabla\tilde\chi)\big],
\end{equation}
\begin{equation}
\label{eq:divergence_nondimensional}
\Dderiv{\tilde\chi}{\tilde t}=\tilde\omega^2-\|\tilde\nabla(\tilde
\nabla^\perp\tilde \psi+\tilde\nabla\tilde\phi)\|_F^2-\frac{4\eta}{
  \Mach^2}\tilde\Delta\zeta+\frac 1\Rey e^{-\zeta}\big[\tilde\Delta
  \tilde\chi-\tilde\nabla\zeta\cdot(\tilde\nabla^\perp\tilde\omega+
  \tilde\nabla\tilde\chi)\big],
\end{equation}
\end{subequations}
while the Navier--Stokes vorticity equation in \eqref{eq:NS_vorticity}
becomes
\begin{equation}
\label{eq:NS_vorticity_nondimensional}
\Dderiv{\tilde\omega_\NS}{\tilde t}=\frac 1\Rey\tilde\Delta\tilde
\omega_\NS.
\end{equation}
Above, the Mach and Reynolds numbers above given, respectively, via
\begin{equation}
\Mach=\Omega L\sqrt{\frac{\rho_0}{p_0}},\qquad\Rey=\frac{\Omega L^2}
\nu.
\end{equation}
Here, observe that the definitions of $\Mach$ and $\Rey$ above are
markedly different from the conventional ones, as the latter make use
of the bulk speed of the flow rather than its vorticity. However, in
the present context, our definitions are more meaningful, because they
are based on the variation of the large scale flow speed inside the
domain, rather than the speed itself (which can be eliminated via a
Galilean shift of the reference frame, as we did above in Section
\ref{sec:background_flow}).

We also note that, at a low Mach number, the coefficient
$4\eta/\Mach^2$ in \eqref{eq:divergence_nondimensional} is not
negligible. In particular, for gases at normal conditions (sea level,
room temperature), the packing fraction $\eta\sim 10^{-3}$
\cite{Abr25}. Thus, setting, for example, $\Mach=0.1$ (which is not a
particularly low value) already leads to $4\eta/\Mach^2=0.4$.

\section{Linearization around the steady state of constant vorticity}
\label{sec:linearization}

To understand the development of generic solutions in a nonlinear
system of equations such as the one in
\eqref{eq:vorticity_divergence_nondimensional}, the standard initial
approach is to linearize this system around its large scale steady
state, and then examine its linearly unstable structures (that is, if
there are any). This can be done with the help of the Fourier
transformation, which converts spatial derivatives into wavevector
multiplications, and the subsequent application of the method of
characteristics.

Here, we note that the linear shear flow in \eqref{eq:linear_shear}
happens to be a steady state for both~\eqref{eq:vorticity_divergence}
and \eqref{eq:NS_vorticity}. Additionally, in the variables of
\eqref{eq:vorticity_divergence}, the linear shear flow
in~\eqref{eq:linear_shear} is a constant vorticity state $(\zeta=0,
\,\omega=\Omega,\,\chi=0)$, which, in the nondimensional variables of
\eqref{eq:vorticity_divergence_nondimensional} and
\eqref{eq:NS_vorticity_nondimensional}, becomes $(\zeta=0,
\,\tilde\omega=1,\,\tilde\chi=0)$.

Mildly abusing the notations, we henceforth re-denote $\tilde\omega$
to be the deviation of the nondimensional vorticity from the unity,
thus making $\tilde\psi$ to refer to the deviation of the
nondimensional streamfunction from its background state $\tilde
y^2/2$. Replacing $\tilde\omega$ and $\tilde\psi$
in~\eqref{eq:vorticity_divergence_nondimensional} with
$\tilde\omega+1$ and $\tilde\psi+\tilde y^2/2$, respectively, and
discarding nonlinear terms, we obtain
\begin{subequations}
\label{eq:linearized_equations}
\begin{equation}
\parderiv\zeta{\tilde t}-\tilde y\parderiv\zeta{\tilde x}=-\tilde\chi,
\qquad\parderiv{\tilde\omega}{\tilde t}-\tilde y\parderiv{\tilde\omega
}{\tilde x}=\frac 1\Rey\tilde\Delta\tilde\omega-\tilde\chi,
\end{equation}
\begin{equation}
\parderiv{\tilde\chi}{\tilde t}-\tilde y\parderiv{\tilde\chi}{\tilde
  x}=\frac 1\Rey\tilde\Delta\tilde\chi+2\left(\parderiv{^2\tilde
  \psi}{\tilde x^2}+\parderiv{^2\tilde\phi}{\tilde x\partial\tilde y}
\right)-\frac{4\eta}{\Mach^2}\tilde\Delta\zeta.
\end{equation}
\end{subequations}
In the Fourier space, \eqref{eq:linearized_equations} becomes a system
of linear transport equations of the first order. Indeed, let us
assume, for convenience, that the spatial domain of
\eqref{eq:linearized_equations} is unbounded, so that the Fourier
transformation in the integral form can be applied. The resulting
Fourier transforms $\hat\zeta(\tilde t,\BV k)$, $\hat\omega(\tilde
t,\BV k)$ and $\hat\chi(\tilde t,\BV k)$, with the wavevector $\BV
k=(k_x,k_y)$, satisfy the following system:
\begin{subequations}
\label{eq:fourier_domain}
\begin{equation}
\parderiv{\hat\zeta}{\tilde t}+k_x\parderiv{\hat\zeta}{k_y}=-\hat\chi,
\qquad\parderiv{\hat\omega}{\tilde t}+k_x\parderiv{\hat\omega}{k_y}
=-\frac{\|\BV k\|^2}\Rey\hat\omega-\hat\chi,
\end{equation}
\begin{equation}
\parderiv{\hat\chi}{\tilde t}+k_x\parderiv{\hat\chi}{k_y}=\bigg(\frac{
  2k_xk_y}{\|\BV k\|^2}-\frac{\|\BV k\|^2}\Rey\bigg)\hat\chi+\frac{4
  \eta\|\BV k\|^2}{\Mach^2}\hat\zeta+\frac{2k_x^2}{\|\BV k\|^2}\hat
\omega.
\end{equation}
\end{subequations}
Also, the linearization of the Navier--Stokes vorticity equation in
\eqref{eq:NS_vorticity_nondimensional} yields
\begin{equation}
\label{eq:NS_linearized}
\parderiv{\tilde\omega_\NS}{\tilde t}-\tilde y\parderiv{\tilde
  \omega_\NS}{\tilde x}=\frac 1\Rey\tilde\Delta\tilde\omega_\NS,
\qquad\parderiv{\hat\omega_\NS}{\tilde t}+k_x\parderiv{\hat\omega_\NS
}{k_y}=-\frac{\|\BV k\|^2}\Rey\hat\omega_\NS.
\end{equation}

\subsection{Transition to ordinary differential equations on
characteristics}

The system of linear equations above in \eqref{eq:fourier_domain}, as
well as the linear equation in \eqref{eq:NS_linearized}, can be
converted into linear ordinary differential equations (ODE) via the
method of characteristics. For that, let us parameterize a
characteristic line, which passes through $k_{y,0}$ at $\tilde t=0$,
via
\begin{equation}
\label{eq:characteristic}
\big(\tilde t,k_y(\tilde t)\big)=(0,k_{y,0})+\tilde t(1,k_x).
\end{equation}
If $f(\tilde t,k_y)$ is a differentiable function of two arguments,
its directional derivative along such a characteristic is
\begin{equation}
\parderiv f{\tilde t}+k_x\parderiv f{k_y}=\deriv{}{\tilde t}f
\big(\tilde t,k_y(\tilde t)\big).
\end{equation}
On this characteristic, the linearized Navier--Stokes vorticity
equation in \eqref{eq:NS_linearized} becomes a single exactly solvable
ODE
\begin{subequations}
\label{eq:NS_ODE}
\begin{equation}
\deriv{\hat\omega_\NS}{\tilde t}=-\frac{\|\BV k(\tilde t)\|^2}\Rey\hat
\omega_\NS,\qquad\hat\omega_\NS(\tilde t)=\hat\omega_\NS(0)\exp\left(
-\frac 1\Rey\int_0^{\tilde t}\|\BV k(r)\|^2\dif r\right),
\end{equation}
\begin{equation}
k_y(\tilde t)=k_x\tilde t+k_{y,0},\qquad\|\BV k(\tilde t)\|^2=k_x^2+
k_y^2(\tilde t).
\end{equation}
\end{subequations}
A generic solution of \eqref{eq:NS_ODE} with $k_x\neq 0$ vanishes as
$\exp(-k_x^2\tilde t^3/3\Rey)$ as $\tilde t\to \infty$, which
corresponds to a viscous decay.

The equations in \eqref{eq:fourier_domain} on a characteristic become
the system of linear ODE of the form
\begin{equation}
\label{eq:fourier_ODE}
\deriv\odeVec{\tilde t}=\BM A(\tilde t)\odeVec,\quad\odeVec=
\begin{pmatrix}\hat\zeta \\ \hat\omega \\ \hat\chi\end{pmatrix},\quad
\BM A(\tilde t)=\begin{pmatrix} 0 & 0 & -1 \\ 0 & -\frac{\|\BV
  k(\tilde t)\|^2}\Rey & -1 \\ \frac{4\eta \|\BV k(\tilde t)\|^2}{
  \Mach^2} & \frac{2k_x^2}{\|\BV k(\tilde t)\|^2} & \frac{2k_xk_y(
  \tilde t)}{\|\BV k(\tilde t)\|^2}-\frac{\|\BV k(\tilde t)\|^2}
\Rey\end{pmatrix}.
\end{equation}
For the further analysis of \eqref{eq:fourier_ODE}, it is convenient
to introduce the following notations:
\begin{subequations}
\label{eq:shorthand}
\begin{equation}
\BV k_0=(k_x,k_{y,0}),\qquad\kappa(\tilde t)=\frac{\|\BV k(\tilde t)\|^2
}{\|\BV k_0\|^2},\qquad\alpha=\frac{\|\BV k_0\|^2}\Rey,
\end{equation}
\begin{equation}
\beta=\frac{\sqrt 2|k_x|}{\|\BV k_0\|},\qquad\varepsilon=\frac\Mach{2
  \sqrt\eta\|\BV k_0\|}=\frac\Mach{2\sqrt{\eta\alpha\Rey}}.
\end{equation}
\end{subequations}
In these notations, $\BM A(\tilde t)$ is written via
\begin{equation}
\label{eq:A_shorthand}
\BM A(\tilde t)=\begin{pmatrix} 0 & 0 & -1 \\ 0 & -\alpha\kappa(\tilde
t) & -1 \\ \kappa(\tilde t)/\varepsilon^2 & \beta^2/\kappa(\tilde t) &
\kappa'(\tilde t)/\kappa(\tilde t)-\alpha\kappa(\tilde t)\end{pmatrix},
\end{equation}
where we note that, coincidentally,
\begin{equation}
\frac{2k_xk_y(\tilde t)}{\|\BV k(\tilde t)\|^2}=\frac{\|\BV k_0\|^2}{
  \|\BV k_0\|^2}\frac{2k_x(k_x\tilde t+k_{y,0})}{(k_x\tilde t+k_{y,0}
  )^2+k_x^2}=\frac{\kappa'(\tilde t)}{\kappa(\tilde t)}.
\end{equation}

\subsection{Phase volume}

The Wronski determinant of \eqref{eq:fourier_ODE} can be computed
exactly:
\begin{equation}
\label{eq:W}
W(\tilde t)=\exp\bigg(\int_0^{\tilde t}\trace\BM A(z)\dif z\bigg)=
\kappa(\tilde t)\exp \left[-\frac {2\tilde t}{3 \Rey}\left(k_x^2\tilde
  t^2+3k_xk_{y,0}\tilde t+3\|\BV k_0\|^2\right)\right].
\end{equation}
The computation is shown in Appendix \ref{sec:app_W}. As we can see,
the phase volume of the system can grow in time in a short term, if
$k_x$ and $k_{y,0}$ are of the same sign. The rate of growth is
maximized when $k_{y,0}=k_x$, that is, when a fluctuation has
comparable scale in both dimensions. As $\tilde t\to\infty$, however,
the phase volume decays as $\sim\exp(-2k_x^2\tilde t^3/3\Rey)$.

\subsection{Eigenvalues and eigenvectors}

The characteristic equation for $\BM A$ in \eqref{eq:A_shorthand} is
\begin{equation}
\label{eq:cubic_equation}
\lambda^3+\left(2\alpha\kappa-\frac{\kappa'}\kappa\right)\lambda^2+
\left((\alpha\kappa)^2- \alpha\kappa'+\frac\kappa{\varepsilon^2}
+\frac{\beta^2}\kappa\right)\lambda+\frac{\alpha\kappa^2}{
  \varepsilon^2}=0.
\end{equation}
This is a cubic equation, and, while technically its roots are
computable explicitly, the corresponding formula for the computation
depends on the sign of the cubic discriminant of
\eqref{eq:cubic_equation}. Besides, even if the sign of the
discriminant is fixed in a given scenario, the resulting expressions,
even if explicit, are too complicated to be meaningful.  Thus, in
order to simplify the expressions for the roots
of~\eqref{eq:cubic_equation}, appropriate estimates will have to be
made. In Appendix~\ref{sec:app_cubic}, we compute approximate
eigenvalues in two different scenarios: in the short-time limit, and
in the asymptotic limit.

For a known eigenvalue $\lambda$ of $\BM A$ in \eqref{eq:A_shorthand},
its corresponding eigenvector $\BV e_\lambda$ is readily
available. Indeed, recall that $\BV e_\lambda$ has one degree of
freedom, and solves $\BM A\BV e_\lambda =\lambda\BV e_\lambda$. Taking
$\hat\chi$ as the necessary degree of freedom, we express, from
\eqref{eq:A_shorthand},
\begin{equation}
\label{eq:eigenvector}
\hat\zeta=-\frac{\hat\chi}\lambda,\qquad\hat\omega=-\frac{\hat\chi}{
 \lambda+\alpha\kappa},\quad\text{and thus}\quad\BV e_\lambda
= \begin{pmatrix} \lambda+\alpha\kappa \\ \lambda
  \\ -\lambda(\lambda+\alpha\kappa)\end{pmatrix}.
\end{equation}

\subsection{Eigenvalues and eigenvectors in the absence of the mean
field potential}

In the absence of the mean field potential, we set
$\varepsilon=\infty$ (or, equivalently, $\eta=0$ in
\eqref{eq:fourier_ODE}). This decouples the $\hat\zeta$-variable from
the $\hat\chi$-equation, which means that it suffices to examine only
the lower-right 2$\times$2 block of $\BM A$, which involves
$\hat\omega$ and $\hat\chi$. For this block, the characteristic
equation is quadratic:
\begin{equation}
\lambda^2+\left(2\alpha\kappa-\frac{\kappa'}\kappa\right)\lambda
+(\alpha\kappa)^2-\alpha\kappa'+\frac{\beta^2}\kappa=0.
\end{equation}
The eigenvalues are given via the complex-conjugate pair
\begin{equation}
\label{eq:eigenvalues_conventional}
\lambda_{1,2}=\frac{\kappa'}{2\kappa}-\alpha\kappa\pm i\sqrt{\frac{
    \beta^2}\kappa-\left(\frac{\kappa'}{2\kappa}\right)^2}=\frac{k_x
  k_y(\tilde t)}{\|\BV k(\tilde t)\|}-\frac{\| \BV k(\tilde t)\|^2}
\Rey\pm\frac{ik_x^2}{\|\BV k(\tilde t)\|^2}\sqrt{2+\frac{k_y^2(\tilde
    t)}{k_x^2}},
\end{equation}
with the corresponding eigenvectors given via
\begin{equation}
\BV e_{1,2}= \begin{pmatrix} 1 \\ -\frac{k_xk_y(\tilde t)}{\|\BV k(\tilde t)\|} \end{pmatrix}\mp\frac{ik_x^2}{\|\BV k(\tilde t)\|^2}\sqrt{2+\frac{k_y^2(\tilde
    t)}{k_x^2}}\begin{pmatrix} 0 \\ 1 \end{pmatrix}.
\end{equation}

\section{Short-term behavior}
\label{sec:short_time}

First, we examine the properties of \eqref{eq:fourier_ODE} at short
times, that is, for $\tilde t\sim 1$.  As we mentioned above,
computing the solution of the characteristic equation
\eqref{eq:cubic_equation} explicitly via cubic formulas leads to
rather complicated expressions, which need to be appropriately
simplified. In the current work, we simplify those expressions by
expanding in power series of a parameter, which can be identified as
either ``large'' or ``small''.

For $\tilde t\sim 1$, we find that the only such parameter is
$\varepsilon$. Indeed, from \eqref{eq:shorthand} observe that
$\kappa\sim 1$, $|\kappa'/\kappa|\sim 1$, while both $\alpha$ and
$\beta$ can be anywhere from zero to $\sim 1$ and $\sqrt 2$,
respectively. At the same time, $\varepsilon\ll 1$, because, as we
mentioned before, at typical low Mach numbers we have
$4\eta/\Mach^2\sim 1$, while it is known from observations that
$\Rey\gg 1$ for turbulent regimes \cite{Rey83}.

Below, in Appendix~\ref{sec:app_cubic_short}, we compute the
approximate roots of \eqref{eq:cubic_equation} using $\varepsilon$ as
a small parameter. The result is
\begin{equation}
\label{eq:eigenvalues_short_time_2}
\lambda_0^s=-\alpha\kappa+O(\varepsilon),\qquad\lambda_{1,2}^s=\frac
12 \left(\frac{\kappa'}\kappa-\alpha\kappa\right)\pm\frac{i\sqrt
  \kappa}\varepsilon+O(\varepsilon).
\end{equation}
The corresponding eigenvectors are computed via
\eqref{eq:eigenvector}:
\begin{equation}
\label{eq:eigenvectors_short_time_2}
\BV e_0^s=\begin{pmatrix} 0 \\ 1 \\ 0\end{pmatrix}+ O(\varepsilon),
\qquad\BV e_{1,2}^s=\begin{pmatrix}\pm\frac{i\varepsilon}{\sqrt\kappa}
\\ \pm\frac{i\varepsilon}{\sqrt\kappa} \\ 1\pm\frac{i\varepsilon}{
  \sqrt\kappa}\frac{\kappa'} \kappa\end{pmatrix}+O(\varepsilon^2).
\end{equation}
For very short times, that is, $\tilde t\ll 1$, the solution of
\eqref{eq:fourier_ODE} is given via $\exp(\tilde t\BM A(0))$ with an
$O(\tilde t^2)$-correction, which can be disregarded if $\tilde t$ is
sufficiently short. Here we note that $\kappa(0)=1$. Reverting
\eqref{eq:eigenvalues_short_time_2} and
\eqref{eq:eigenvectors_short_time_2} back to the notations of
\eqref{eq:fourier_ODE}, we have
\begin{subequations}
\label{eq:eigenvalues_short_time}
\begin{equation}
\lambda_0^s=-\frac{\|\BV k_0\|^2}\Rey+O(\Rey^{-1/2}),\qquad\BV e_0^s
=\begin{pmatrix} 0 \\ 1 \\ 0\end{pmatrix}+O(\Rey^{-1/2}),
\end{equation}
\begin{equation}
\lambda_{1,2}^s=\frac{k_xk_{y,0}}{\|\BV k_0\|^2}-\frac{\|\BV k_0\|^2}{
  2\Rey} \pm\frac{2i\sqrt\eta\|\BV k_0\|}\Mach+O(\Rey^{-1/2}),\qquad
\BV e_{1,2}^s=\begin{pmatrix} 0 \\ 0 \\ 1\end{pmatrix}+O(\Rey^{-1/2}).
\end{equation}
\end{subequations}
Assuming that $O(\Rey^{-1/2})$-terms can be disregarded, observe that
$\lambda_0<0$, which means that the direction corresponding to
$\hat\omega$ is stable. At the same time, the real part of the
complex-conjugate pair $\lambda_{1,2}$ can be positive, with the
necessary condition that $k_x$ and $k_{y,0}$ have matching signs.
Particularly, choosing $k_{y,0}=k_x$ maximizes the real part of
$\lambda_{1,2}$:
\begin{equation}
\lambda_{1,2}^{s,\text{max}}=\frac 12\left(1-\frac{\|\BV k_0\|^2}
\Rey\right)\pm\frac{2i\sqrt\eta\|\BV k_0\|}\Mach+ O(\Rey^{-1/2}).
\end{equation}
As we can see, the eigenvalues are unstable for $\|\BV k_0\|^2<\Rey$.
For these linearly unstable wavenumbers, the rate of exponential
growth is $O(1)$, however, the frequency of oscillations, induced by
the imaginary parts, is $O(\|\BV k_0\|)$, that is, inversely
proportional to the spatial scale of a fluctuation.  These rapid
oscillations at small scales are caused by the coupling of the mean
field potential with the large scale background vorticity, and
manifest primarily in the direction of $\hat\chi$-variable, that is,
the divergence of the flow velocity. Also, the fact that $k_{y,0}=k_x$
sets the fastest local growth of fluctuations, suggests that the most
``visible'' fluctuations are comparable in scale in both dimensions
(which seems to match observations).

In the $(k_x,k_y)$-plane, the characteristics in
\eqref{eq:characteristic} are straight vertical lines, parallel to the
$k_y$-axis. The direction of movement along a characteristic is upward
in the right-hand half of the plane (that is, for $k_x>0$), and
downward in the left-hand half of the plane. At the same time, the
linearly unstable regions, given via $\Re(\lambda_{1,2}^s)>0$, are
located in the first and third quadrants of the plane, where $k_x$ and
$k_y$ have matching signs.  Therefore, those solutions, which are
amplified by the linearly unstable regions, may not cross the
horizontal axis of the plane (that is, $k_y=0$), because they move
further toward higher wavenumbers, or smaller scales (the so-called
``direct cascade'').  This observation becomes important below in
Section~\ref{sec:power_scaling}, where we discuss the Kolmogorov
energy scaling.

At the same time, in the absence of the mean field potential, we
obtain, from \eqref{eq:eigenvalues_conventional},
\begin{equation}
\lambda_{1,2}=\frac{k_xk_{y,0}}{\|\BV k_0\|^2}-\frac{\|\BV k_0\|^2}
\Rey\pm\frac{ik_x^2}{\|\BV k_0\|^2}\sqrt{2+\left(\frac{k_{y,0}}{k_x}
  \right)^2}.
\end{equation}
Here, the linearly unstable wavenumbers extend only up to $\|\BV
k_0\|^2<\Rey/2$. Also, it is easy to see that the imaginary part of
$\lambda_{1,2}$ cannot exceed $\sqrt 2$, which means that the
frequency of induced oscillations at linearly unstable wavenumbers is
slow (that is, no more than $O(1)$) across all spatial scales.

\section{Asymptotic behavior}
\label{sec:long_time}

As $\tilde t\to\infty$, it is clear that $\kappa\sim\tilde
t^2\to\infty$, while $\alpha$, $\beta$ and $\varepsilon$ remain
fixed. Therefore, below in Appendix~\ref{sec:app_cubic_long} we compute
the roots of the characteristic equation \eqref{eq:cubic_equation}
using $\kappa^{-1}$ as a small parameter, which yields
\begin{equation}
\label{eq:eigenvalues_long_time}
\lambda_0^l=-\frac 1{\alpha\varepsilon^2}+O(\kappa^{-1}),\qquad
\lambda_1^l=-\alpha\kappa+\frac 1{\alpha\varepsilon^2}+\frac{\kappa'}
\kappa+O(\kappa^{-1}),\qquad\lambda_2^l=-\alpha\kappa+O(\kappa^{-1}).
\end{equation}
The corresponding eigenvectors are computed via
\eqref{eq:eigenvector}:
\begin{equation}
\BV e_0^l=\begin{pmatrix} \alpha\varepsilon^2 \\ 0 \\ 1 \end{pmatrix}
+O(\kappa^{-1}),\qquad\BV e_1^l=\begin{pmatrix} 0 \\ -\alpha\varepsilon^2
\\ 1+\alpha\varepsilon^2\kappa'/\kappa \end{pmatrix}+O(\kappa^{-1}),
\qquad\BV e_2^l = \begin{pmatrix} 0 \\ 1 \\ 0 \end{pmatrix}+O(\kappa^{-1}).
\end{equation}
Reverting back to the notations of \eqref{eq:fourier_ODE}, we obtain
\begin{subequations}
\label{eq:eigenvalues_eigenvectors_long_time}
\begin{equation}
\label{eq:lambda0_long_time}
\lambda_0^l=-\frac{4\eta\Rey}{\Mach^2}+O(\tilde t^{-2}),\qquad\BV
e_0^l =\begin{pmatrix} \frac{\Mach^2}{4\eta\Rey} \\ 0 \\ 1
\end{pmatrix}+O(\tilde t^{-2}),
\end{equation}
\begin{equation}
\lambda_1^l=-\frac{\|\BV k(\tilde t)\|^2}{\Rey}+\frac{4\eta\Rey}{\Mach^2}
+\frac{2k_xk_y(\tilde t)}{\|\BV k(\tilde t)\|^2}+O(\tilde t^{-2}),\qquad\BV e_1^l
=\begin{pmatrix} 0 \\ -\frac{\Mach^2}{4\eta\Rey} \\ 1+\frac{\Mach^2}{2\eta\Rey}
\frac{k_xk_y(\tilde t)}{\|\BV k(\tilde t)\|^2}\end{pmatrix}+O(\tilde t^{-2}),
\end{equation}
\begin{equation}
\lambda_2^l =-\frac{\|\BV k(\tilde t)\|^2}\Rey+O(\tilde
t^{-2}),\qquad\BV e_2^l = \begin{pmatrix} 0 \\ 1
  \\ 0 \end{pmatrix}+O(\tilde t^{-2}).
\end{equation}
\end{subequations}
Below in Appendix \ref{sec:app_stability}, we show that
\eqref{eq:fourier_ODE} is asymptotically stable as $\tilde
t\to\infty$. Moreover, a fundamental solution of
\eqref{eq:fourier_ODE} asymptotically behaves as
\begin{equation}
\label{eq:asymptotic_solution}
\odeVec(\tilde t)\sim C_0e^{-\frac{4\eta\Rey}{\Mach^2}\tilde t}
\begin{pmatrix} \frac{\Mach^2}{4\eta\Rey} \\ 0 \\ 1\end{pmatrix}+
  C_1e^{-\frac{k_x^2}{3\Rey}\tilde t^3}\begin{pmatrix} 0 \\ -\frac{\Mach^2}{4
 \eta\Rey} \\ 1\end{pmatrix}+C_2e^{-\frac{k_x^2}{3\Rey}\tilde t^3}
    \begin{pmatrix} 0 \\ 1 \\ 0 \end{pmatrix}.
\end{equation}
For a very large $\tilde t$, the first term above dominates the
remaining two, that is,
\begin{equation}
\label{eq:persistent_solution}
\odeVec(\tilde t)\sim e^{-\frac{4\eta\Rey}{\Mach^2}\tilde t}
\begin{pmatrix}\frac{\Mach^2}{4\eta\Rey} \\ 0 \\ 1\end{pmatrix},
\end{equation}
which means that $\odeVec(\tilde t)$ decays with a constant
exponential rate across all scales (i.e., the asymptotic decay in
\eqref{eq:fourier_ODE} is not viscous, unlike that in the
incompressible vorticity dynamics \eqref{eq:NS_ODE}). In the context
of \eqref{eq:fourier_domain}, these solutions are traveling waves
which exponentially decay in time at a constant rate. This is a rather
peculiar phenomenon, because all explicit damping terms in
\eqref{eq:fourier_domain} are viscous, that is, the decay rate they
confer is scale-selective.

In the absence of the mean field potential, in Appendix
\ref{sec:app_stability_conventional} we show that
\begin{equation}
\label{eq:asymptotic_conventional}
\begin{pmatrix}\hat\omega \\ \hat\chi\end{pmatrix}\sim
 e^{-\frac{k_x^2}{3\Rey}\tilde t^3},
\end{equation}
that is, the dynamics for the vorticity and the velocity divergence
are asymptotically stable with a viscous decay.

\subsection{Apparent universality of the decay rate in a persistent
solution}

It is interesting that the decay rate in
\eqref{eq:persistent_solution} is independent of the spatial scale of
the flow. Indeed, observe that the quotient $\eta\Rey/\Mach^2$ is
given via
\begin{equation}
\frac{\eta\Rey}{\Mach^2}=\frac{\rho_0}{\rho_\HS}\frac{\Omega L^2}\nu
\frac{p_0}{\rho_0\Omega^2L^2}=\frac{p_0}{\rho_\HS\Omega\nu},
\end{equation}
and thus depends only on the hard sphere density $\rho_\HS$, the
viscosity $\nu$, the background pressure $p_0$, and the large scale
vorticity $\Omega$. Moreover, assuming that the large scale vorticity
$\Omega$ is the result of an external effect acting on the flow (such
as the Coriolis force, for example), which is generally damped by the
viscosity $\nu$, it is possible that, in practice, the product $\Omega
\nu$ itself is largely a function of the strength of such an external
effect. If the strength of the external effect, together with the
background pressure of the flow, is roughly constant at a given
location (for instance, at a particular latitude), the whole
coefficient may appear to be a universal constant to an observer.

\section{A hypothesis on the Kolmogorov energy scaling}
\label{sec:power_scaling}

Above, we found that the asymptotically persistent solution in
\eqref{eq:persistent_solution} is aligned with the eigenvector $\BV
e_0^l$ in \eqref{eq:lambda0_long_time} as $\tilde t\to\infty$.  Note
that, for $\Rey\gg 1$, $\BV e_0^l$ is primarily aligned with
$\hat\chi$.  Here, we demonstrate that, in the long time limit, the
kinetic energy is produced by this persistent velocity divergence
$\hat\chi$, and suggest a hypothesis why its $k_x$-scaling is
proportional to $k_x^{-5/3}$ \cite{Kol41a,Kol41c}, as observed in
nature and experiments \cite{BucVel,NasGag}.

First, we take the velocity equation from
\eqref{eq:inertial_flow_zeta} and apply the same nondimensionalization
as in~\eqref{eq:vorticity_divergence_nondimensional}, while expressing
the advection operator in terms of the partial derivatives:
\begin{equation}
\parderiv{\BV{\tilde u}}{\tilde t}+(\BV{\tilde u}\cdot\tilde\nabla)
\BV{\tilde u}+\frac{4\eta}{\Mach^2}\tilde\nabla\zeta=\frac 1\Rey
e^{-\zeta}\tilde\Delta\BV{\tilde u},\qquad\BV u=\Omega L\BV{\tilde u}.
\end{equation}
Next, we introduce the energy tensor $\BM E=\BV{\tilde u}^2/2$, with
the streamwise component in the upper-left corner denoted via
$E_{xx}=u_x^2/2$.  Expressing the advection term as
\begin{equation}
(\BV{\tilde u}\cdot\tilde\nabla)\BV{\tilde u}=\tilde\nabla\cdot(\BV{
    \tilde u}^2)-(\tilde\nabla\cdot\BV{\tilde u})\BV{\tilde u}=2\tilde
  \nabla\cdot\BM E-(\tilde\nabla \cdot\BV{\tilde u})\BV{\tilde u},
\end{equation}
and recalling that $\tilde\nabla \cdot\BV{\tilde u}=\tilde\chi$, we
arrive at
\begin{equation}
\parderiv{\BV{\tilde u}}{\tilde t}+2\tilde\nabla\cdot\BM E-\tilde\chi
\BV{\tilde u}+\frac{4\eta}{\Mach^2}\tilde\nabla\zeta=\frac 1\Rey
e^{-\zeta}\tilde\Delta\BV{\tilde u}.
\end{equation}
Next, we apply the divergence operator to the whole equation above,
\begin{equation}
\parderiv{\tilde\chi}{\tilde t}+2\tilde\nabla^2:\BM E-\nabla\tilde
\chi\cdot\BV{\tilde u}-\tilde\chi^2+\frac{4\eta}{\Mach^2}\tilde\Delta
\zeta=\frac 1\Rey\tilde\nabla\cdot(e^{-\zeta}\tilde\Delta\BV{\tilde
  u}),
\end{equation}
where ``$:$'' refers to the Frobenius product of two matrices.
Finally, we apply the same linearization as in
\eqref{eq:linearized_equations}, which yields $\BV{\tilde u}_0=
(-\tilde y,0)^T$, and leads to
\begin{equation}
\nabla\tilde\chi\cdot\BV{\tilde u}\to\nabla\tilde\chi\cdot\BV{\tilde
  u}_0=-\tilde y\parderiv{\tilde\chi}{ \tilde x},\qquad\tilde\chi^2\to
0,\qquad\tilde\nabla\cdot(e^{-\zeta} \tilde\Delta\BV{\tilde
  u})\to\tilde\Delta\tilde\chi.
\end{equation}
The resulting linearized equation and its Fourier transform are,
respectively,
\begin{subequations}
\begin{equation}
\parderiv{\tilde\chi}{\tilde t}+\tilde y\parderiv{\tilde\chi}{\tilde
  x}+2\tilde\nabla^2:\BM E+\frac{4\eta}{\Mach^2}\tilde\Delta\zeta=
\frac 1\Rey\tilde\Delta\tilde\chi,\quad\text{and}
\end{equation}
\begin{equation}
\label{eq:energy_fourier}
\parderiv{\hat\chi}{\tilde t}-k_x\parderiv{\hat\chi}{k_y}-2\BV k^2:\hE
=\frac{4\eta\|\BV k\|^2}{\Mach^2}\hat\zeta-\frac{\|\BV k\|^2}\Rey\hat
\chi.
\end{equation}
\end{subequations}
Let us rearrange \eqref{eq:energy_fourier} as
\begin{equation}
\parderiv{\hat\chi}{\tilde t}-\frac 12\bigg[\bigg(\parderiv{}{\tilde
    t}+k_x\parderiv{}{k_y}\bigg)\hat\chi+\bigg(\frac{4\eta\hat\zeta}{
    \Mach^2}-\frac{\hat\chi}\Rey\bigg)\|\BV k\|^2\bigg]=\BV k^2:\hE.
\end{equation}
If the solution along a characteristic is of the form
\eqref{eq:persistent_solution}, then, first, the multiple of $\|\BV
k\|^2$ in the left-hand side vanishes as $\tilde t\to\infty$, and,
second, the directional derivative is equivalent to the multiplication
by $\lambda_0^l$ from \eqref{eq:lambda0_long_time}. Therefore, we
conclude that, for a sufficiently large $\tilde t$, the equation above
simplifies to
\begin{equation}
\label{eq:E}
\left(\parderiv{}{\tilde t}+\frac{2\eta\Rey}{\Mach^2}\right)\hat\chi
=\BV k^2:\hE,\qquad\text{or}\qquad\left(\parderiv{}{\tilde t}+\frac{2
  \eta\Rey}{\Mach^2}\right)\hat\chi=k_x^2\hat E_{xx}\quad\text{at}
\quad k_y=0.
\end{equation}
We note that, in the physical space, $k_y=0$ corresponds to an average
in the $\tilde y$-direction (that is, across the direction of the
large scale flow).

\subsection{Asymptotic scaling of persistent solutions}

The next step is to determine whether there might exist a
$k_x$-scaling of $\partial_{\tilde t} \hat\chi$. The complication here
lies primarily in the fact that the partial time differentiation is
transversal to the direction of characteristics, and, therefore, in
order to estimate it, we need to have understanding of the shape of
the solution across characteristics. To this end, we assume that a
persistent solution $\odeVec(\tilde t,k_y)$ of \eqref{eq:fourier_ODE},
for some large $\tilde t$, originates from an initial state
$\odeVec(\tilde t_0,k_{y,0})$, for $\tilde t_0\sim 1$. Denoting the
principal matrix of \eqref{eq:fourier_ODE} via $\BM P(\tilde t,\tilde
t_0)$, we write
\begin{equation}
\label{eq:ODE_solution}
\odeVec(\tilde t,k_y)=\BM P(\tilde t,\tilde t_0)\odeVec(\tilde
t_0,k_{y,0}).
\end{equation}
Since the characteristic, along which $\BM P(\tilde t,\tilde t_0)$
propagates, passes through $(\tilde t,k_y)$ and $(\tilde t_0,
k_{y,0})$, and its slope is given via $k_x$, we can express
$k_{y,0}=k_y-k_x(\tilde t-\tilde t_0)$:
\begin{equation}
\odeVec(\tilde t,k_y)=\BM P(\tilde t,\tilde t_0)\odeVec(\tilde
t_0,k_y-k_x(\tilde t-\tilde t_0)).
\end{equation}
Further, we express $t_0=t-T$, where $T$ is the elapsed time between
$t_0$ and $t$:
\begin{equation}
\odeVec(\tilde t,k_y)=\BM P(\tilde t,\tilde t-T)\odeVec(\tilde t-T,k_y-k_x
T).
\end{equation}
From the way the above expression is arranged, it is clear that the
partial differentiation in $\tilde t$ in the left-hand side leads to
the differentiation of the ``initial condition'' $\odeVec(\tilde
t_0,k_{y,0})$ with respect to its first argument. Therefore, in order
to proceed, we need to make a reasonable assumption about the
dependence of $\odeVec(\tilde t_0,k_{y,0})$ on $t_0$.

Here, we assume that $\odeVec(\tilde t_0,k_{y,0})$ itself is produced
by the short-time linearly unstable eigenvalues in
\eqref{eq:eigenvalues_short_time} from small random uncorrelated
fluctuations (which, in turn, means that $\tilde t_0\sim 1$, and
$k_{y,0}\sim k_x$). Therefore, along the $k_y$-axis (that is, for a
fixed $\tilde t_0$), $\odeVec(\tilde t_0,k_{y,0})$ must be described
via a generic function $\BV F(-k_{y,0})$, which by itself does not
have an inherent $k_x$-scaling. Flipping $\BV F$ along the $\tilde
t$-axis (that is, fixing $k_{y,0}$ and varying $\tilde t_0$), and
noting that the slope of the characteristic is $k_x$, we have
$\odeVec(\tilde t_0,k_{y,0})=\BV F(k_x\tilde t_0)$.  Thus, we can
express
\begin{equation}
\odeVec(\tilde t,k_y)=\BM P(\tilde t,\tilde t-T)\BV F(k_x(\tilde t-T)).
\end{equation}
As we can see, partial differentiation in $\tilde t$ involves not only
the derivative of $\BV F$ (whose dependence on $k_x$ is provided), but
also the derivatives of the principal matrix $\BM P(\tilde t,\tilde
t_0)$ of \eqref{eq:fourier_ODE} in both of its arguments.

To obtain a crude estimate for the $k_x$-scaling of $\partial_{\tilde
  t}\odeVec(\tilde t,k_y)$ without having to differentiate the
principal matrix, we resort to the following approximation. Observe
that rescaling the time as $s\sim\tilde t^3$ in \eqref{eq:fourier_ODE}
leads to an asymptotically autonomous linear system (for more details,
see Appendix~\ref{sec:app_stability}). Here, we denote the principal
matrix of the rescaled system as $\BM P_*(s,s_0)\equiv\BM P(\sqrt
[3]s,\sqrt[3]{s_0})$, and write, with the help of the cocycle
property of $\BM P$,
\begin{equation}
\label{eq:X_t_s}
\odeVec(\tilde t,k_y)=\BM P\left(\tilde t,\sqrt[3]s\right)\BM P_*(s,s_0)
\BV F\left(k_x\sqrt[3]{s_0}\right).
\end{equation}
Above, $\tilde t$, $s$ and $s_0$ are independent parameters, although
we assume that $s_0\sim 1$, $\tilde t\gg 1$, and $s\sim\tilde t^3$.
Next, just like for $\tilde t$ and $\tilde t_0$ above, we express
$s_0=s-S$, which leads to
\begin{equation}
\odeVec(\tilde t,k_y)=\BM P\left(\tilde t,\sqrt[3]s\right)\BM P_*(s,s-S)
\BV F\left(k_x\sqrt[3]{s-S}\right).
\end{equation}
Next, we assume that, for our choice of $s$ and $s_0$, $\BM
P_*(s,s_0)=\BM P_*(s-s_0)$, i.e. the principal matrix of the
time-rescaled system is autonomous. This further leads to
\begin{equation}
\odeVec(\tilde t,k_y)=\BM P\left(\tilde t,\sqrt[3]s\right)\BM P_*(S)\BV
F\left(k_x\sqrt[3]{s-S}\right).
\end{equation}
In what follows, we set $S$ to a constant, so that $\BM P_*(S)$
becomes a constant matrix. In order to be able to compute the
derivative in $\tilde t$, it remains to set $s=\tilde t^3$. Since $\BM
P(\tilde t,\tilde t)$ is the identity matrix, we arrive at
\begin{equation}
\odeVec(\tilde t,k_y)=\BM P_*(S)\BV F\left(k_x\sqrt[3]{\tilde t^3-S}
\right).
\end{equation}
Next, we recall that $\sqrt[3]{\tilde t^3-S}=\sqrt[3]{s_0}=\tilde
t_0$. We henceforth fix $\tilde t_0$ as a constant, and use a small
parameter $\delta$ to describe the variation of the expression under
the cubic root:
\begin{subequations}
\begin{equation}
\tilde t^3-S=\tilde t_0^3(1+\delta),\qquad\delta=\frac{\tilde t^3-S}{
  \tilde t_0^3}-1,
\end{equation}
\begin{equation}
\sqrt[3]{\tilde t_0^3 (1+ \delta)}=\tilde t_0\left(1+\frac\delta 3+
O(\delta^2)\right)=\frac{\tilde t^3}{3\tilde t_0^2} +\frac{\tilde t_0
}3\bigg(2-\frac S{\tilde t_0^3}\bigg)+O(\delta^2).
\end{equation}
\end{subequations}
Therefore, for those $\tilde t$ for which $\tilde t^3-S$ is close
enough to $\tilde t_0^3$, and $O(\delta^2)$ can be discarded, we can
express, approximately,
\begin{equation}
\odeVec(\tilde t,k_y)=\BM P_*(S)\BV F\left(\frac{k_x\tilde t^3}{3
  \tilde t_0^2}+\frac{k_x\tilde t_0}3\bigg(2-\frac S{\tilde t_0^3}
\bigg) \right).
\end{equation}
The partial differentiation in $\tilde t$ then yields
\begin{multline}
\label{eq:chain_rule}
\parderiv{}{\tilde t}\odeVec(\tilde t,k_y)=\BM P_*(S)\frac{k_x\tilde
  t^2}{\tilde t_0^2}\BV F'\left(\frac{k_x\tilde t^3}{3\tilde t_0^2}
+\frac{k_x\tilde t_0}3\bigg(2-\frac S{\tilde t_0^3}\bigg)\right)\\=
\BM P(\tilde t,\tilde t_0)\sqrt[3]{\frac{k_x}{3\tilde t_0^2}}
\parderiv{}z\BV F\left(z^3-c(\tilde t^3) \right),\qquad z=\tilde
t\sqrt[3]{\frac{k_x}{3\tilde t_0^2}},\qquad c(r)=k_x\tilde t_0
\bigg(\frac r{3\tilde t_0^3}-1\bigg),
\end{multline}
where we reverse-engineered the chain rule in the second line and
replaced $S=\tilde t^3-\tilde t_0^3$. Finally, observe that, for large
$\tilde t$, $\BM P(\tilde t,\tilde t_0)$ is a projection matrix onto
$\BV e_0^l$ in \eqref{eq:lambda0_long_time}, whose eigenvalue does not
depend on $k_x$, and therefore, $\BM P$ itself does not scale with
$k_x$. It becomes clear that $\partial_{\tilde t}\odeVec(\tilde
t,k_y)\sim\sqrt[3]{k_x}$, and, in particular, $\partial_{\tilde
  t}\hat\chi(\tilde t,k_y)\sim\sqrt[3]{k_x}$.

\subsection{The necessity for the inverse cascade and its possible
mechanics}

Substituting~\eqref{eq:ODE_solution} and \eqref{eq:chain_rule}
into~\eqref{eq:E}, and denoting the projection $\hat\chi(\tilde t)=\BM
P_{\hat\chi}(\tilde t,\tilde t_0)\odeVec_0$ we arrive at
\begin{equation}
\BV k^2:\hE(\tilde t,k_y)=\BM P_{\hat\chi}(\tilde t,\tilde t_0)\left[
  \sqrt[3] {\frac{k_x}{3\tilde t_0^2}}\parderiv{}z\BV F\left(z^3-
  c(\tilde t^3)\right)+\frac{2\eta\Rey}{\Mach^2}\BV F(k_x\tilde
  t_0)\right].
\end{equation}
If the first term in parentheses dominates, then the right-hand side
scales as $\sqrt[3]{k_x}$. Further, if $k_y=0$, then $\BV k^2:
\hE(\tilde t,0)=k_x^2\hat E_{xx}(\tilde t,0)$, and, therefore, $\hat
E_{xx}\sim |k_x|^{-5/3}$, which is the Kolmogorov scaling
\cite{Kol41a,Kol41c}. Conversely, if the second term dominates, then
$\hat E_{xx}\sim |k_x|^{-2}$.

However, note that any persistent solution, which crosses the axis
$k_y=0$ in the $(k_x,k_y)$-plane, arrives from linearly stable
quadrants (that is, the second and the fourth), and constitutes a
so-called ``inverse cascade'' -- that is, the corresponding spatial
pattern starts at small scales and expands into large scales in the
$\tilde y$-variable (that is, across the direction of the large scale
flow). As we remarked above in Section~\ref{sec:linearization}, it is
impossible to create the inverse cascade within the context of the
linearized problem we examine here. Particularly, those persistent
solutions, which are created from small fluctuations in the linearly
unstable first and third quadrants of the $(k_x,k_y)$-plane,
inevitably escape into small scales (the direct cascade).

Therefore, in order to receive persistent solutions from the linearly
stable second and fourth quadrants, we need at least two such
linearized problems cross-feeding their inverse cascades with locally
generated persistent solutions. This appears to be possible in the
case of two adjacent regions, extending along the direction of the
flow, where the large scale vorticity has opposite signs (the simplest
example is a straight jet). In the geometry of a straight jet, it
appears that two such regions should be able to ``communicate''
between each other across the direction of the large scale flow, such
that those persistent solutions, which are created in one region and
escape into the small scales of its first or third quadrant, emerge in
the second or fourth quadrant of the other region at the small scales,
thus feeding its inverse cascade. This, however, appears to be a much
more complicated problem, and is clearly beyond the scope of the
current work.

\subsection{Relation to observations and experiments}

Lastly, we relate the computation of the Kolmogorov energy scaling to
the results of observations and experiments. In particular, from the
foregoing, we summarize that:
\begin{enumerate}[label=\arabic*.]
\item The Kolmogorov energy scaling appears to be produced by
  persistent solutions of the velocity divergence of
  \eqref{eq:linearized_equations}, through the interaction of at least
  two regions of large scale vorticity of opposite signs;
\item Such persistent solutions develop from small random fluctuations
  of the velocity divergence in the originating region by the
  unstable, rapidly oscillating eigenvalues in
  \eqref{eq:eigenvalues_short_time};
\item These persistent solutions escape into the direct cascade of the
  originating region, and should manifest at the small scales of the
  complementary region;
\item The Kolmogorov energy scaling seems to be produced in the
  complementary region when the persistent solutions reach large
  scales via the inverse cascade.
\end{enumerate}
Therefore, it appears that, in order for the Kolmogorov energy scaling
to be produced and observed, the following conditions should generally
be fulfilled:
\begin{enumerate}[label={\alph*)}]
\item The background flow has two adjacent regions of vorticity of
  opposite signs;
\item The persistent solutions do not have intrinsic $k_x$-scaling;
\item The measurements correspond to $k_y=0$ (an average over the
  $\tilde y$-coordinate).
\end{enumerate}
Usually, all three conditions hold in typical experimental or
observational settings. First, the background flow is normally a jet,
which indeed has two regions of vorticity of opposite signs extending
symmetrically along its axis (for a graphical representation, refer,
for example, to Figure~2 in our work \cite{Abr23}). Second, the real
parts of the unstable eigenvalues in~\eqref{eq:eigenvalues_short_time}
clearly lack discernible power scaling in $k_x$ (they, of course,
depend on $k_x$, but not in a manner which would be visible on a
log-log plot as a slanted straight line). Third, the measurements are
usually statistical averages of the streamwise kinetic energy of the
flow, which do not depend on the transversal coordinate (see
\cite{BucVel} or \cite{NasGag} as examples).

Finally, we have to note that the foregoing cannot happen in the
absence of the mean field potential, because the solutions decay at a
viscous, scale-selective rate.

\section{Discussion}
\label{sec:discussion}

In our recent works \cite{Abr22,Abr23,Abr24,Abr26}, we found that the
presence of the mean field effect of an intermolecular potential in
the equations for inertial flow results in a spontaneous development
of turbulent motions in an initially laminar flow. In particular, in
our work~\cite{Abr24} we also carried out a benchmark simulation
without the mean field effect, and observed that the resulting
numerical solution remained laminar.

In the current work, we examine the two-dimensional inertial flow
equations, linearized around a background state of constant vorticity,
in order to understand how the presence of the mean field potential
affects solutions, and to identify a possible cause of the spontaneous
development of turbulent motions. We also make a comparison to
similarly linearized incompressible Navier--Stokes equations. The
summary of results follows below.

First, we find that, while there are no instabilities in solutions of
the linearized incompressible Navier--Stokes equations (that is, their
solutions decay to zero monotonically), the inertial flow equations
possess regions of linearly unstable wavenumbers at short time scales,
which exponentially amplify initial conditions. While there is an
increase in overall linear instability of the laminar steady state of
the inertial flow due to the presence of the mean field potential, it
is largely insignificant. However, what we also discover is that, at
the linearly unstable wavenumbers, the coupling of the mean field
potential with the background vorticity induces rapid oscillations of
fluctuations of the velocity divergence. The frequency of these
oscillations scales inversely proportionally to the spatial size of
the fluctuation, so that larger fluctuations oscillate more slowly,
while smaller fluctuations oscillate faster. We find that, at a given
wavenumber, the most unstable wavevector has equal components, i.e.
the fastest growing fluctuations are of comparable sizes in both
dimensions. In the absence of the mean field potential, the imaginary
parts of the unstable eigenvalues do not scale with the wavenumber,
and thus do not confer rapid oscillations to the solution at small
scales.

At long time scales, we find an interesting eigenvector, which is also
aligned largely with the divergence of velocity, and which allows
asymptotically stable, but persistent solutions to propagate along
characteristics in the Fourier space in the form of decaying traveling
waves. Remarkably, these solutions decay at a constant exponential
rate, which does not depend on the wavenumber, and even does not
depend on the overall spatial scale of the flow, being only a function
of the viscosity, large scale vorticity, and background pressure.
This finding is especially intriguing because there are only viscous,
scale-selective explicit damping terms in the inertial flow equations,
and nothing overtly indicates the existence of solutions with a
uniform, constant decay rate at all scales. Further, it appears that
the Kolmogorov scaling of the kinetic energy of the flow is produced
precisely by this persistent divergence of velocity of the flow. The
negative five-thirds power of the wavenumber in the Kolmogorov scaling
seems to emerge due to the cubic time dilation, which makes the
solutions along characteristics autonomous. At the same time, in the
absence of the mean field potential, solutions become asymptotically
stable with the conventional, viscous, scale-selective rate of decay.

In the course of nondimensionalization of the inertial flow equations,
we express the Mach and Reynolds numbers using the large scale
vorticity, rather than the speed of the flow, as one of the reference
constants. From the standpoint of physics, our interpretation is more
meaningful, because it is based on the overall variation of the flow
speed over the domain, rather than the speed by itself -- the latter
can be changed via a Galilean shift of the frame of reference. In
addition, the experiments, where the flow remains laminar at a high
Reynolds number (see, for example, \cite{Pfe}), seem to be designed
purposely to reduce the vorticity of the flow, as well as to better
isolate the flow from possible external fluctuations of the divergence
of velocity (such as randomly passing acoustic waves). In such an
experiment, our vorticity-based definition of the Reynolds number
would likely yield a more realistic value of the latter.

\ack The reasoning behind \eqref{eq:X_t_s}--\eqref{eq:chain_rule} is
owed to Oliver the Poodle and his retractable leash. The work was
supported by the Simons Foundation grant \#636144.

\appendix

\section{Computation of the Wronski determinant}
\label{sec:app_W}

The Wronski determinant of \eqref{eq:fourier_ODE} is given via
\begin{subequations}
\begin{equation}
W(\tilde t)=\exp\bigg(\int_0^{\tilde t}\trace\BM A(r)\dif r\bigg)
=\exp\left[2\int_0^{\tilde t}\left( \frac{k_xk_y(r)}{\|\BV k(r)\|^2}
  -\frac{\|\BV k(r)\|^2}\Rey\right)\dif r\right],
\end{equation}
\begin{multline}
-\frac 2\Rey\int_0^{\tilde t}\|\BV k(r)\|^2\dif r=-\frac {2k_x^2}\Rey
\int_0^{\tilde t}\left[(r+k_{y,0}/k_x)^2+1\right]\dif r\\=-\frac{2k_x^2
}{3\Rey}\left[(\tilde t+k_{y,0}/k_x)^3-(k_{y,0}/k_x)^3+3\tilde t\right]
=-\frac {2\tilde t}{3 \Rey}\left(k_x^2\tilde t^2+3k_xk_{y,0}\tilde t+
3(k_x^2 +k_{y,0}^2)\right),
\end{multline}
\begin{equation}
\int_0^{\tilde t}\frac{2k_xk_y(r)}{\|\BV k(r)\|^2}\dif r=\int_0^{
  \tilde t}\frac{2(r+k_{y,0}/k_x)}{(r+k_{y,0}/k_x)^2+1}\dif r=\ln
\bigg(\frac{k_x^2+(k_x\tilde t+k_{y,0})^2}{k_x^2+k_{y,0}^2}\bigg),
\end{equation}
\end{subequations}
which leads to \eqref{eq:W}.

\section{Computation of eigenvalues}
\label{sec:app_cubic}

The roots of \eqref{eq:cubic_equation} are computed as follows. First,
the reduction to a depressed cubic equation is made via the
substitution
\begin{equation}
\label{eq:lambda_gamma}
\gamma=\frac\lambda\kappa+\frac 13\left(2\alpha-\frac{\kappa'}{
  \kappa^2}\right).
\end{equation}
The result is
\begin{subequations}
\begin{equation}
\gamma^3+p\gamma+q=0,\qquad p=\frac 1{\varepsilon^2\kappa}+\frac{
  \beta^2}{\kappa^3}-\frac 14\frac{(\kappa')^2}{\kappa^4}-\frac 13
\left(\alpha-\frac 12\frac{\kappa' }{\kappa^2}\right)^2,
\end{equation}
\begin{equation}
q=\frac 1{3\varepsilon^2 \kappa}\left(\alpha+ \frac{\kappa'}{\kappa^2}
\right)-\frac 1{27}\left(2\alpha-\frac{\kappa'}{\kappa^2}\right)
\left(\alpha^2-\alpha\frac{\kappa' }{\kappa^2}-\frac{2(\kappa')^2}{
  \kappa^4}+\frac{9\beta^2}{\kappa^3}\right).
\end{equation}
\end{subequations}
The sign of the cubic discriminant, given via
\begin{equation}
D=\left(\frac q2\right)^2+\left(\frac p3\right)^3,
\end{equation}
determines which formula to use to compute the roots.

The expressions for $p$ and $q$ are complicated, and thus we will use
estimates. According to the assumptions in
Section~\ref{sec:short_time}, $\alpha\sim 1$, $\beta\sim 1$,
$\varepsilon\ll 1$. However, $\kappa$ varies between unity (for short
times) and infinity (for long times). Thus, we have to make estimates
for the short-time and long-time scenarios separately.

\subsection{Short-time scenario}
\label{sec:app_cubic_short}

Here, we treat $\varepsilon$ as a small parameter, and subsequently
expand the calculations in powers of $\varepsilon$. In this case, we
can estimate $p\sim 1/\varepsilon^2\kappa>0$, and Cardano's formula
applies. According to Cardano's formula, the roots are given via
\begin{equation}
\gamma_0=\xi_--\xi_+,\qquad\gamma_{1,2}=\frac{1\pm i\sqrt 3}2\xi_+
-\frac{1\mp i\sqrt 3}2\xi_-,\qquad\xi_\pm=\left(\sqrt D\pm \frac
q2\right)^{1/3}.
\end{equation}
Using the fact that $\varepsilon$ is a small parameter, we express
\begin{subequations}
\begin{equation}
p^3=\frac 1{(\varepsilon^2\kappa)^3}(1+O(\varepsilon^2)),\qquad q^2=
O(\varepsilon^{-4}),\qquad D=\frac 1{(3\varepsilon^2\kappa)^3} (1+
O(\varepsilon^2)),
\end{equation}
\begin{equation}
\sqrt{D}=\frac 1{(3\kappa)^{3/2}\varepsilon^3}(1+O(\varepsilon^2)),
\qquad q=\frac 1{3\varepsilon^2 \kappa}\left(\alpha+
\frac{\kappa'}{\kappa^2} \right)+O(1),
\end{equation}
\begin{equation}
\sqrt D\pm \frac q2=\frac 1{(3\kappa)^{3/2}\varepsilon^3}\left(1\pm
\frac{\varepsilon\sqrt{3\kappa}}2\left(\alpha+\frac{\kappa'}{\kappa^2}
\right)+O(\varepsilon^2)\right),
\end{equation}
\begin{equation}
\xi_\pm=\left(\sqrt D\pm \frac q2\right)^{1/3}=\frac 1{\varepsilon
  \sqrt{3\kappa}}\pm \frac 16\left(\alpha+\frac{\kappa'}{\kappa^2}
\right)+O(\varepsilon),
\end{equation}
\begin{equation}
\gamma_0=-\frac 13\left(\alpha+\frac{\kappa'}{\kappa^2}\right)+
O(\varepsilon),\qquad\gamma_{1,2}=\frac 16\left(\alpha+\frac{\kappa'}{
  \kappa^2}\right)\pm \frac i{\varepsilon\sqrt\kappa} +O(\varepsilon),
\end{equation}
\end{subequations}
from which \eqref{eq:eigenvalues_short_time_2} follows via the reverse
application of \eqref{eq:lambda_gamma}.

\subsection{Long-time scenario}
\label{sec:app_cubic_long}

Here, we treat $\kappa^{-1}$ as a small parameter, and expand the
calculations in negative powers of $\kappa$, noting that
$|\kappa'/\kappa|=O(\kappa^{-1/2})$:
\begin{subequations}
\begin{equation}  
p=-\frac{\alpha^2}3+\frac 1\kappa\left(\frac 1{\varepsilon^2}+\frac
\alpha 3\frac{\kappa'}\kappa\right)+O(\kappa^{-3}),\quad q=-\frac{2
  \alpha^3}{27}+\frac\alpha{3\kappa}\left(\frac 1{\varepsilon^2} +
\frac\alpha 3\frac{\kappa'}\kappa\right)+\frac{\kappa'}{3
  \varepsilon^2\kappa^3}+O(\kappa^{-3}),
\end{equation}
\begin{multline}
D=\left[\frac{\alpha^3}{27}-\frac\alpha{6\kappa}\left(\frac 1{
    \varepsilon^2}+\frac\alpha 3\frac{\kappa'}\kappa\right)+\frac{
    \kappa' }{6\varepsilon^2\kappa^3}+O(\kappa^{-3})\right]^2-\left[
  \frac{\alpha^2}9-\frac 1{3\kappa}\left(\frac 1{\varepsilon^2}+\frac
  \alpha 3\frac{\kappa'}\kappa\right)+O(\kappa^{-3})\right]^3
\\=-\frac{\alpha^2}{18\varepsilon^2 \kappa^2}\left(\frac
1{2\varepsilon^2}+\frac{\alpha\kappa'}{9\kappa}
\right)+O(\kappa^{-3}).
\end{multline}
\end{subequations}
Since $D<0$ for sufficiently large $\kappa$, there are three real
roots, and, therefore, the trigonometric formula applies for
$j=0,1,2$:
\begin{multline}
\gamma_j=\frac 2{\sqrt 3}\left(\frac{\alpha^2}3-\frac 1\kappa\left(
\frac 1{\varepsilon^2}+\frac\alpha 3\frac{\kappa'}\kappa\right)+
O(\kappa^{-3})\right)^{1/2}\\\cos\Bigg\{\frac 13\arccos\Bigg[\frac{3
    \sqrt 3}{2}\left(\frac{2 \alpha^3}{27}-\frac\alpha{3\kappa}\left(
  \frac 1{\varepsilon^2}+\frac\alpha 3\frac{\kappa'}\kappa\right)-
  \frac{\kappa'}{3 \varepsilon^2\kappa^3}+O(\kappa^{-3})\right)
  \\ \left(\frac{\alpha^2}3-\frac 1\kappa\left(\frac 1{\varepsilon^2}
  +\frac\alpha 3\frac{\kappa'}\kappa\right)+O(\kappa^{-3}) \right)
  ^{-3/2}\Bigg]-\frac 23\pi j\Bigg\}=\left(\frac{2\alpha}3-\frac
1{\alpha\kappa}\left( \frac 1{\varepsilon^2}+\frac\alpha 3\frac{
  \kappa'}\kappa\right)+ O(\kappa^{-2})\right)\\\cos\left[\frac 13
  \arccos\left( 1-\frac{27}{8\alpha^4\varepsilon^4\kappa^2}\left(1+
  \frac{2\alpha\varepsilon^2\kappa'}\kappa\right)+O(\kappa^{-3})
  \right)-\frac 23\pi j\right].
\end{multline}
We recall that a power expansion of $\arccos(1-x)$ is
\begin{equation}
\arccos(1-x)=\sqrt{2x}+\frac{(2x)^{3/2}}{24}+O(x^2),\qquad\text{for }
x\geq 0,
\end{equation}
which yields
\begin{equation}
\arccos\left(1-\frac{27}{8\alpha^4\varepsilon^4\kappa^2}\left(1+
\frac{2\alpha\varepsilon^2\kappa'}\kappa\right)+O(\kappa^{-3})\right)
=\frac{3\sqrt 3}{2\alpha^2\varepsilon^2\kappa}\left(1+\frac{\alpha
  \varepsilon^2\kappa'}\kappa\right)+O(\kappa^{-2}).
\end{equation}
Therefore, the expression for $\gamma_j$ now becomes
\begin{equation}
\gamma_j=\left(\frac{2\alpha}3-\frac 1{\alpha\kappa}\left(\frac 1{
  \varepsilon^2}+\frac\alpha 3\frac{\kappa'}\kappa\right)+
O(\kappa^{-2})\right)\cos\left(-\frac 23\pi j+\frac{\sqrt 3}{
  2\alpha^2\varepsilon^2\kappa}\left(1+\frac{\alpha\varepsilon^2
  \kappa'}\kappa\right)+O(\kappa^{-2})\right).
\end{equation}
In particular, for $\gamma_0$ we have
\begin{multline}
\gamma_0=\left(\frac{2\alpha}3-\frac 1{\alpha\kappa}\left(\frac 1{
  \varepsilon^2}+\frac\alpha 3\frac{\kappa'}\kappa\right)+O(\kappa^{
  -2})\right)\cos\left(\frac{\sqrt 3}{2\alpha^2\varepsilon^2\kappa}
\left(1+\frac{\alpha\varepsilon^2\kappa'}\kappa\right)+O(\kappa^{-2})
\right)\\ =\frac{2\alpha}3-\frac 1{\alpha\kappa}\left(\frac 1{
  \varepsilon^2}+\frac\alpha 3\frac{\kappa'}\kappa\right)+
O(\kappa^{-2}).
\end{multline}
For $j=1,2$, we observe that
\begin{equation}
\cos\left(-\frac{2\pi j}3+x\right)=\cos\left(-\frac{2\pi j}3\right)
+\cos'\left(-\frac{2\pi j}3\right)x+O(x^2)=-\frac 12\pm\frac{\sqrt
  3}2x+O(x^2),
\end{equation}
which yields
\begin{multline}
\gamma_{1,2}=\left(\frac{2\alpha}3-\frac 1{\alpha\kappa}\left(\frac 1{
  \varepsilon^2}+\frac\alpha 3\frac{\kappa'}\kappa\right)+
O(\kappa^{-2})\right)\left(-\frac 12\pm\frac 3{
  4\alpha^2\varepsilon^2\kappa}\left(1+\frac{\alpha\varepsilon^2
  \kappa'}\kappa\right)+O(\kappa^{-2})\right)\\= -\frac\alpha 3+\frac
1{2\alpha\varepsilon^2\kappa}+\frac{
  \kappa'}{6\kappa^2}\pm\frac 1{ 2\alpha\varepsilon^2 \kappa}
\pm\frac{\kappa'}{2\kappa^2}+O(\kappa^{-2}).
\end{multline}
Assembling the results above together, we obtain
\begin{subequations}
\begin{equation}
\gamma_0=\frac{2\alpha}3-\frac 1{\alpha\kappa}\left(\frac 1{
  \varepsilon^2}+\frac\alpha 3\frac{\kappa'}\kappa\right)+
O(\kappa^{-2}),
\end{equation}
\begin{equation}
\gamma_1=-\frac\alpha 3+\frac 1{\alpha\varepsilon^2\kappa}+\frac{2
  \kappa'}{3\kappa^2}+O(\kappa^{-2}),\qquad\gamma_2=-\frac\alpha
3-\frac{\kappa'}{3\kappa^2}+O(\kappa^{-2}),
\end{equation}
\end{subequations}
from which \eqref{eq:eigenvalues_long_time} follows via the reverse
application of \eqref{eq:lambda_gamma}.

\section{Asymptotic behavior}
\label{sec:app_stability}

Here, we examine the asymptotic behavior of \eqref{eq:fourier_ODE}
with the help of Levinson's theorem. To this end, we introduce a
suitable change of the time variable. Namely, we define the new time
variable $s$ via
\begin{equation}
\label{eq:tau}
s=\tau(\tilde t)=\tilde t+\tilde t^3/3,\qquad\tau'(\tilde t)=1+\tilde
t^2.
\end{equation}
Note that $\tau'(\tilde t)>0$ for all $\tilde t$, and thus
$\tau(\tilde t)$ is monotonically increasing and invertible on the
whole real line, that is, $\tilde t=\tau^{-1}(s)$. In order to express
\eqref{eq:fourier_ODE} in the new time variable, we divide the whole
system by $\tau'$, and note that $\tau'(\tilde t)\dif{\tilde t}=\dif
s$. The result is the following system of linear ordinary differential
equations:
\begin{equation}
\label{eq:fourier_ODE_rescaled}
\deriv\odeVec s=\BM B(s)\odeVec,\qquad\BM B(s)=\frac{\BM A}{\tau'}=
\begin{pmatrix} 0 & 0 & -1/\tau' \\ 0 & -\alpha\kappa/\tau'
& -1/\tau' \\ \kappa/\varepsilon^2\tau' & \beta^2/\kappa\tau' &
(\kappa'/\kappa-\alpha\kappa)/\tau'
\end{pmatrix},
\end{equation}
where $\kappa=\kappa(\tau^{-1}(s))$, $\tau'=\tau'(\tau^{-1}(s))$. For
a large $s$, $\kappa\sim\tau'\sim s^{2/3}$. Clearly, the eigenvectors
of $\BM B$ are those of $\BM A$ in
\eqref{eq:eigenvalues_eigenvectors_long_time}, while the eigenvalues
of $\BM B$ are those of $\BM A$ in
\eqref{eq:eigenvalues_eigenvectors_long_time}, further divided by
$\tau'$:
\begin{subequations}
\label{eq:eigenvalues_B}
\begin{equation}
\lambda_0^B(s)=-\frac 1{\alpha\varepsilon^2\tau'}+O(s^{-4/3}),
\qquad\BV e_0^l(s)=\begin{pmatrix} \alpha\varepsilon^2 \\ 0 \\ 1 \end{pmatrix}
+O(s^{-2/3})
\end{equation}
\begin{equation}
\lambda_1^B(s)=-\frac{\alpha\kappa}{\tau'}+\frac 1{\alpha
  \varepsilon^2\tau'}+\frac{\kappa'}{\kappa\tau'}+O(s^{-4/3}),
\qquad\BV e_1^l(s)=\begin{pmatrix} 0 \\ -\alpha\varepsilon^2
\\ 1+\alpha\varepsilon^2\kappa'/\kappa \end{pmatrix}+O(s^{-2/3}),
\end{equation}
\begin{equation}
\lambda_2^B(s)=-\frac{\alpha\kappa}{\tau'}+O(s^{-4/3}),\qquad
\BV e_2^l(s) = \begin{pmatrix} 0 \\ 1 \\ 0 \end{pmatrix}+O(s^{-2/3}).
\end{equation}
\end{subequations}
As $s\to\infty$, $\BM B$ in \eqref{eq:fourier_ODE_rescaled} becomes a
constant matrix with finite entries. Indeed, observe that all nonzero
entries of $\BM B$ decay to zero, with the exception of those which
involve the quotient $\kappa/\tau'$.  Moreover, the quotient
$\kappa/\tau'$ above in \eqref{eq:fourier_ODE_rescaled} is strictly
positive, bounded and continuously differentiable on a real line as
long as $k_x\neq 0$ (the relevant calculations are shown in
Appendix~\ref{sec:app_kappatau}).  Therefore, the matrix $\BM B$ in
\eqref{eq:fourier_ODE_rescaled} is bounded and continuously
differentiable on the whole real line, and, for any initial state
$(s_0,\hat\zeta_0,\hat\omega_0,\hat\chi_0)$, the solution is unique
and exists on the whole real line. In turn, this means that the whole
space of $(s,\hat\zeta,\hat\omega,\hat\chi)$ is densely filled with
the integral curves of~\eqref{eq:fourier_ODE_rescaled}. Additionally,
we have $\kappa/\tau'\to k_x^2/\|\BV k_0\|^2$ as $s\to\infty$, which
leads to
\begin{equation}
\BM B(\infty)=\frac{k_x^2}{\|\BV k_0\|^2}
\begin{pmatrix} 0 & 0 & 0 \\ 0 & -\alpha
& 0 \\ \varepsilon^{-2} & 0 & -\alpha
\end{pmatrix}.
\end{equation}
From \eqref{eq:eigenvalues_B}, observe that, for a finite $s$, $\BM
B(s)$ has three distinct real eigenvalues with the corresponding
linearly independent eigenvectors. Yet, as $s\to\infty$, the two lower
eigenvalues in $\BM B(\infty)$ coalesce into $-k_x^2/\Rey$, with any
vector from the $\hat\omega\hat\chi$-plane being an eigenvector. To
ameliorate this quirk, we note that the geometric and algebraic
multiplicities of the lower eigenvalue of $\BM B(\infty)$ match, and
the vectors $\BV e_1^l(\infty)$ and $\BV e_2^l(\infty)$
from~\eqref{eq:eigenvalues_B} can be chosen as the pair of linearly
independent eigenvectors for it, since both lie in the
$\hat\omega\hat\chi$-plane. With this choice, $\BM B(s)\to\BM
B(\infty)$ continuously with all its eigenvalues and eigenvectors, and
thus is continuously diagonalizable for all $s$ as $s\to\infty$.

Next, we denote $\BM L(s)=\diag\big(\lambda_0^B(s),\lambda_1^B(s),
\lambda_2^B(s)\big)$, $\BM E(s)=\big(\BV e_0^l(s),\BV e_1^l(s),\BV
e_2^l(s)\big)$, such that $\BM B(s)=\BM E(s)\BM L(s)\BM E^{-1}(s)$.
Denoting $\odeVec(s)=\BM E(s)\BV Z(s)$, and observing
that
\begin{equation}
\deriv{\BM E^{-1}(s)}s=-\BM E^{-1}(s)\BM E'(s)\BM E^{-1}(s),
\end{equation}
we write~\eqref{eq:fourier_ODE_rescaled} in the form
\begin{equation}
\BV Z'(s)=\left(\BM L(s)+\BM E^{-1}(s)\BM E'(s)\right)\BV Z(s).
\end{equation}
Now, we have
\begin{enumerate}[label={\alph*)}]
\item $\|\BM E'(s)\|\sim s^{-4/3}$, and thus $\displaystyle\int_s^\infty\|\BM
  E^{-1}(r)\BM E'(r)\|\dif r<\infty$,
\item $\lambda_0^B(s)=O(s^{-2/3})<0$, and thus $\displaystyle\int_s^\infty
  \lambda_0^B(r)\dif r=-\infty$,
\item $\lambda_1^B-\lambda_0^B=O(1)<0$, and thus $\displaystyle
  \int_s^\infty (\lambda_1^B(r)-\lambda_0^B(r))\dif r=-\infty$,
\item $\lambda_2^B-\lambda_1^B=O(s^{-2/3})<0$, and thus $\displaystyle
  \int_s^\infty (\lambda_2^B(r)-\lambda_1^B(r))\dif r=-\infty$.
\end{enumerate}
Therefore, the integrability and spectral gap conditions of Levinson's
theorem (see \cite{CodLev}, Chap.~3, the first part of Theorem 8.1,
pp.~93--95 up to the Lemma) apply to $\BM E^{-1}(s)\BM E'(s)$ and $\BM
L(s)$, respectively, which leads to the asymptotic estimate for a
fundamental solution
\begin{equation}
\BV Z(s)\sim C_0e^{\int_{s_0}^s\lambda_0^B(r)\dif r}\begin{pmatrix}
  1 \\ 0 \\ 0\end{pmatrix}+C_1e^{\int_{s_0}^s\lambda_1^B(r)\dif r}
  \begin{pmatrix} 0 \\ 1 \\ 0\end{pmatrix}+C_2e^{\int_{s_0}^s
   \lambda_0^B(r)\dif r}\begin{pmatrix} 0 \\ 0 \\ 1\end{pmatrix},
\end{equation}
and, subsequently,
\begin{equation}
\odeVec(s)\sim C_0e^{\int_{s_0}^s\lambda_0^B(r)\dif r}\BV e_0^l(\infty)
  +C_1e^{\int_{s_0}^s\lambda_1^B(r)\dif r}\BV e_1^l(\infty)+C_2e^{\int_{s_0}^s
   \lambda_0^B(r)\dif r}\BV e_2^l(\infty).
\end{equation}
Substituting the leading order expressions
from~\eqref{eq:eigenvalues_B}, computing the integrals, factoring the
terms with $s_0$ into constants, and replacing $s\to\tilde t^3/3$
yields~\eqref{eq:asymptotic_solution}.

As all three eigenvalues integrate to $-\infty$ as $s\to\infty$, the
fundamental solution above decays to zero, and, therefore,
\eqref{eq:fourier_ODE_rescaled} is asymptotically stable. As
\eqref{eq:fourier_ODE_rescaled} follows from \eqref{eq:fourier_ODE}
via a time rescaling, the latter is also asymptotically stable.

\subsection{Asymptotic behavior in the absence of the mean field potential}
\label{sec:app_stability_conventional}

Here, we denote the lower-right 2$\times$2 block of $\BM B(s)$ in
\eqref{eq:fourier_ODE_rescaled} as $\BM B_2(s)$. We further split it
as
\begin{equation}
\BM B_2(s)=-\frac{k_x^2}\Rey\BM I+\BM C(s),
\end{equation}
with $\|\BM C(s)\|\to 0$ as $s\to\infty$. Then, by Corollary 3.21 from
\cite{Tes},
\begin{equation}
\|\odeVec_2(s)\|\sim e^{-\frac{k_x^2}\Rey s},\qquad\text{where}\quad\BV
\odeVec_2 =\begin{pmatrix}\hat\omega \\ \hat\chi\end{pmatrix}.
\end{equation}
Substituting $s\to\tilde t^3/3$
yields~\eqref{eq:asymptotic_conventional}.

\section{Computation of the quotient \texorpdfstring{$\kappa/\tau'$}{k/t'}}
\label{sec:app_kappatau}

From \eqref{eq:shorthand} and \eqref{eq:tau}, we note that the
quotient $\kappa/\tau'$ in \eqref{eq:fourier_ODE_rescaled} is given
via
\begin{equation}
\label{eq:app_kappatau}
\frac\kappa{\tau'}=\frac 1{1+\tilde t^2}\frac{k_x^2(1+\tilde t^2)+2k_x
  k_{y,0}\tilde t+k_{y,0}^2}{k_x^2+k_{y,0}^2}=\frac 1{1+K^2}\left(1+K
\frac{2\tilde t+K}{1+\tilde t^2}\right),\qquad K=\frac{k_{y,0}}{k_x},
\end{equation}
and approaches $(1+K^2)^{-1}$ at $\tilde t\to\pm\infty$.  If we
restrict ourselves to $\tilde t\geq 0$, then, for a given value of
$K\in\RR$, the extremum of $\kappa/\tau'$ (which is either a maximum,
or a minimum, depending on the sign of $K$) is computed by equating
its derivative to zero, resulting in
\begin{subequations}
\begin{equation}
\tilde t^2+K\tilde t-1=0,\qquad\tilde t_{\mathrm{min}/\mathrm{max}}
=\frac{\sqrt{K^2+4}-K}2,
\end{equation}
\begin{equation}
\label{eq:app_kt}
\left.\frac\kappa{\tau'}\right|_{\mathrm{min}/\mathrm{max}}=\frac 1{1+
  K^2}\frac{\sqrt{K^2+4}+K}{\sqrt{K^2+4 }-K}.
\end{equation}
\end{subequations}
We can see that, as $\tilde t$ starts at zero and increases to
$+\infty$, $\kappa/\tau'$ starts at 1, then either increases to its
maximum (if $K>0$) or decreases to its minimum (if $K<0$), and, after
that, asymptotically approaches $(1+K^2)^{-1}$. The absolute maximum
of $\kappa/\tau'$ can also be computed explicitly by equating the
derivative of \eqref{eq:app_kt} to zero, and is given via
\begin{equation}
\max_{\tilde t\geq 0,\;K\in\RR}\frac\kappa{\tau'}=\frac 34\qquad
\text{at}\qquad(\tilde t,K)=\left(\frac 1{\sqrt 2},\frac 1{\sqrt
  2}\right).
\end{equation}
Note that, from \eqref{eq:app_kappatau}, extending $\tilde t$ to
negative values is equivalent to flipping the sign of $K$, so
everything above reverses signs in appropriate places, with the global
maximum of $\kappa/\tau'$ remaining $3/4$ (albeit for $(\tilde t, K)$
with opposite signs).


\begin{thebibliography}{10}

\bibitem{Rey83}
O.~Reynolds.
\newblock {III}. {A}n experimental investigation of the circumstances which
  determine whether the motion of water shall be direct or sinuous, and of the
  law of resistance in parallel channels.
\newblock \href{https://doi.org/10.1098/rspl.1883.0018}{{\em Proc. R. Soc.
  Lond.}}, 35(224--226):84--99, 1883.

\bibitem{Cho}
A.J. Chorin.
\newblock Vorticity and turbulence.
\newblock In J.E. Marsden and L.~Sirovich, editors, {\em Applied Mathematical
  Sciences}, volume 103. Springer, New York, 1994.

\bibitem{Fri}
U.~Frisch.
\newblock {\em Turbulence: The Legacy of {A.N. K}olmogorov}.
\newblock Cambridge University Press, Cambridge, UK, 1995.

\bibitem{MajBer}
A.J. Majda and A.L. Bertozzi.
\newblock {\em Vorticity and Incompressible Flow}.
\newblock Cambridge Texts in Applied Mathematics. Cambridge University Press,
  2002.

\bibitem{AviMoxLozAviBarHof}
K.~Avila, D.~Moxey, A.~de~Lozar, M.~Avila, D.~Barkley, and B.~Hof.
\newblock The onset of turbulence in pipe flow.
\newblock \href{https://doi.org/10.1126/science.1203223}{{\em Science}},
  333:192--196, 2011.

\bibitem{BarSonMukLemAviHof}
D.~Barkley, B.~Song, V.~Mukund, G.~Lemoult, M.~Avila, and B.~Hof.
\newblock The rise of fully turbulent flow.
\newblock \href{https://doi.org/10.1038/nature15701}{{\em Nature}},
  526:550--553, 2015.

\bibitem{KhaAnwHasSan}
H.H. Khan, S.F. Anwer, N.~Hasan, and S.~Sanghi.
\newblock Laminar to turbulent transition in a finite length square duct
  subjected to inlet disturbance.
\newblock \href{https://doi.org/10.1063/5.0048876}{{\em Phys. Fluids}},
  33:065128, 2021.

\bibitem{Vel}
A.~Vela-Mart\'in.
\newblock The energy cascade as the origin of intense events in small-scale
  turbulence.
\newblock \href{https://doi.org/10.1017/jfm.2022.117}{{\em J. Fluid Mech.}},
  937:A13, 2022.

\bibitem{Abr22}
R.V. Abramov.
\newblock Macroscopic turbulent flow via hard sphere potential.
\newblock \href{https://doi.org/10.1063/5.0060121}{{\em AIP Adv.}},
  11(8):085210, 2021.

\bibitem{Abr23}
R.V. Abramov.
\newblock Turbulence in large-scale two-dimensional balanced hard sphere gas
  flow.
\newblock \href{https://doi.org/10.3390/atmos12111520}{{\em Atmosphere}},
  12(11):1520, 2021.

\bibitem{Abr24}
R.V. Abramov.
\newblock Creation of turbulence in polyatomic gas flow via an intermolecular
  potential.
\newblock \href{https://doi.org/10.1103/PhysRevFluids.7.054605}{{\em Phys. Rev.
  Fluids}}, 7(5):054605, 2022.

\bibitem{Abr26}
R.V. Abramov.
\newblock Turbulence via intermolecular potential: Viscosity and transition
  range of the {R}eynolds number.
\newblock \href{https://doi.org/10.3390/fluids8030101}{{\em Fluids}}, 8(3):101,
  2023.

\bibitem{Abr25}
R.V. Abramov.
\newblock Turbulence via intermolecular potential: A weakly compressible model
  of gas flow at low {M}ach number.
\newblock \href{https://doi.org/10.1063/5.0128281}{{\em Phys. Fluids}},
  34(12):125104, 2022.

\bibitem{Gra}
H.~Grad.
\newblock On the kinetic theory of rarefied gases.
\newblock \href{https://doi.org/10.1002/cpa.3160020403}{{\em Comm. Pure Appl.
  Math.}}, 2(4):331--407, 1949.

\bibitem{Cha}
J.G. Charney.
\newblock Geostrophic turbulence.
\newblock
  \href{https://doi.org/10.1175/1520-0469(1971)028<1087:GT>2.0.CO;2}{{\em J.
  Atmos. Sci.}}, 28(6):1087--1095, 1971.

\bibitem{Kol41a}
A.N. Kolmogorov.
\newblock The local structure of turbulence in incompressible viscous fluid for
  very large {R}eynolds numbers.
\newblock \href{https://doi.org/10.1098/rspa.1991.0075}{{\em Dokl. Akad. Nauk
  SSSR}}, 30:299--303, 1941.

\bibitem{Kol41c}
A.N. Kolmogorov.
\newblock Dissipation of energy in the locally isotropic turbulence.
\newblock \href{https://doi.org/10.1098/rspa.1991.0076}{{\em Dokl. Akad. Nauk
  SSSR}}, 32:19--21, 1941.

\bibitem{BucVel}
P.~Buchhave and C.M. Velte.
\newblock Measurement of turbulent spatial structure and kinetic energy
  spectrum by exact temporal-to-spatial mapping.
\newblock \href{https://doi.org/10.1063/1.4999102}{{\em Phys. Fluids}},
  29(8):085109, 2017.

\bibitem{NasGag}
G.D. Nastrom and K.S. Gage.
\newblock A climatology of atmospheric wavenumber spectra of wind and
  temperature observed by commercial aircraft.
\newblock
  \href{https://doi.org/10.1175/1520-0469(1985)042<0950:ACOAWS>2.0.CO;2}{{\em
  J. Atmos. Sci.}}, 42(9):950--960, 1985.

\bibitem{Pfe}
W.~Pfenninger.
\newblock Boundary layer suction experiments with laminar flow at high
  {R}eynolds numbers in the inlet length of a tube by various suction methods.
\newblock In G.V. Lachmann, editor, {\em Boundary Layer and Flow Control},
  pages 961--980. Pergamon, Oxford, UK, 1961.

\bibitem{CodLev}
E.A. Coddington and N.~Levinson.
\newblock {\em Theory of Ordinary Differential Equations}.
\newblock Robert E. Krieger Publishing Company, Inc., Malabar, FL 32950, 1984.

\bibitem{Tes}
G.~Teschl.
\newblock {\em Ordinary Differential Equations and Dynamical Systems}, volume
  140 of {\em Graduate Studies in Mathematics}.
\newblock American Mathematical Society, 2012.

\end{thebibliography}
\end{document}